\documentclass[manuscript=false,nonacm=true]{acmart} 
\usepackage{textcomp}
\usepackage{tikz}
\usepackage{natbib}
\usetikzlibrary{shapes.geometric}
\AtBeginDocument{%
  }

\usepackage{xpatch}

\makeatletter
\xpatchcmd{\ps@firstpagestyle}{Manuscript submitted to ACM}{}{\typeout{First patch succeeded}}{\typeout{first patch failed}}
\xpatchcmd{\ps@standardpagestyle}{Manuscript submitted to ACM}{}{\typeout{Second patch succeeded}}{\typeout{Second patch failed}}    \@ACM@manuscriptfalse
\makeatother
\renewcommand\footnotetextcopyrightpermission[1]{} 

\setcopyright{none}
\copyrightyear{2024}
\acmYear{2024}
\acmDOI{}

\definecolor{reviseorange}{RGB}{255, 165, 0}

\newenvironment{revise}
  {\color{reviseorange}}  
  {}                      

\newcommand*\E{\ensuremath\mathbb{E}}
\newcommand*\DKL{\ensuremath D_{\text{KL}}}

\begin{document}

\title{Active Inference and Human--Computer Interaction}

\author{Roderick Murray-Smith}
\email{Roderick.Murray-Smith@glasgow.ac.uk}
\orcid{0000-0003-4228-7962}
\author{John H. Williamson}
\orcid{0000-0001-8085-7853}
\email{JohnH.Williamson@glasgow.ac.uk}
\author{Sebastian Stein}

\affiliation{%
  \institution{School of Computing Science, University of Glasgow}
  \city{Glasgow}
  \country{Scotland}}

\renewcommand{\shortauthors}{Murray-Smith, Williamson, Stein}






\begin{abstract}
{\it Active Inference} is a closed-loop computational theoretical basis for understanding behaviour, based on agents with internal  probabilistic generative models that encode their beliefs about how hidden states in their environment cause their sensations. We review Active Inference and how it could be applied to model the human-computer interaction loop. Active Inference provides a coherent framework for managing generative models of humans, their environments, sensors and interface components. It informs off-line design and supports real-time, online adaptation. It provides model-based explanations for behaviours observed in HCI, and new tools to measure important concepts such as agency and engagement.  We discuss how Active Inference offers a new basis for a theory of interaction in HCI, tools for design of modern, complex sensor-based systems, and integration of artificial intelligence technologies, enabling it to cope with diversity in human users and contexts. We discuss the practical challenges in implementing such Active Inference-based systems.

\end{abstract}

\maketitle

\section{Introduction}
This paper will examine the incorporation of {\it Active Inference} in the theory and practice of human--computer interaction. Active Inference (AIF) is a closed-loop computational theory for modelling  agent behaviour \cite{friston2006free,ParPezFri22}. AIF assumes a (biological or artificial) agent has an internal generative, probabilistic model which encodes beliefs about how hidden states in the environment cause its sensations, and that it acts to reduce uncertainty to minimise `expected surprise'. Such agents establish and preserve favourable homeostasis by minimising surprise over all predicted futures \cite{friston2011action}. 

Active Inference has become the source of intense scientific interest across a broad set of areas, including consciousness research, philosophy and neuroscience \cite{clark2015surfing,seth2021being,clark2023experience,bogacz2017tutorial}. We intend this paper to open up links between HCI and this frontier of theories. We argue that there is a pressing need for a broad-ranging and rigorous theory for interaction design and analysis; one which we believe Active Inference affords us. \citet{hornbaek2017interaction} highlight: {\it ``While we have concepts of interaction that talk about design ..., we have failed to produce theories and concepts that have both high determinacy and adequate scope.''} and go on to discuss that classical mathematical and simulation models of users often have high determinacy but inadequate scope.


Active Inference is an elegant, unifying theory which incorporates machine learning, Bayesian inference, probabilistic programming, and dynamic systems, and can be applied to design decisions from low-level physical interactions to high-level reasoning and decision tasks. It is a precise, mathematical theory that leads to implementable algorithms, but its scope ranges from the chemotaxis of single-celled organisms to human psychology.  It is an exciting scientific paradigm, and the implications for interaction are profound and stimulating. It is also a theory which many find notoriously difficult to understand \cite{alexander2018, ajmaren2017} and carries deeper significance than it might initially appear.\footnote{Despite the similar name, Active Inference is not directly related to \textit{Active Learning} in the machine learning literature.}

Active Inference provides a basis for scientific, quantitative analysis and engineering of interactive systems which furnishes detailed predictions at a range of levels and with broad application. This paper will present a first proposal for the integration of Active Inference theory into interaction design, viewing the interaction loop as a negotiated, mutual control process between human and computer, with both acting to control their perceptions to minimise their sensory surprise. It will explore the potential impact on HCI theory for off-line analysis, as well as on-line use, creating interactive systems that are better aligned with users' preferences, and can adapt to their needs, ability and context more rapidly.  


\subsection{Contributions}

\begin{figure}[htb]
    \includegraphics[width=\linewidth]{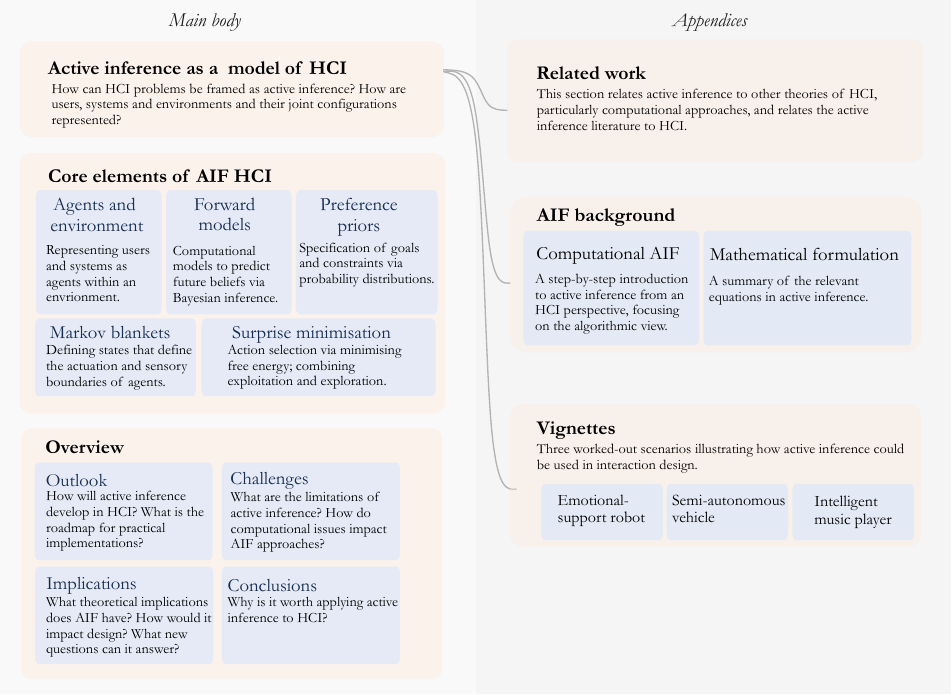}
    \caption{The overall structure of the paper.}
\end{figure}

This is a theory paper without implementations, evaluations or results. Our contributions are:
\begin{enumerate}
    \item An introduction to Active Inference theory for an HCI audience (§\ref{sec:active}), where a tutorial on the algorithms (§\ref{sec:algorithm}) and mathematical detail (§\ref{sec:maths}) is limited to the Appendix.  This includes a review of the AIF literature and how Active Inference relates to other theories of interaction in (§\ref{sec:related}) and through illustrative scenarios (§\ref{sec:vignettes});
    \item An analysis of how AIF principles may be applied in interaction, presenting the general configurations of AIF models in the interaction loop (§\ref{sec:ainf_hci});
    \item A discussion of the hallmark characteristics of AIF theory in interaction, including interaction with {\it Environments, Markov Blankets}, {\it Forward Models} (§\ref{sec:distinctive}) and the role of {\it Prediction}, and representing an agent's goals or desires via {\it Preference priors};
    \item A discussion of the ramifications of AIF approaches. The theory offers novel perspectives into questions of agency, freedom and resilience as well as practical techniques for interaction engineering (§\ref{sec:benefits}), but it is subject to challenges in modelling, computational issues and the limited development of software (§\ref{sec:challenges}).
\end{enumerate}

\section{Active Inference}\label{sec:active}

\subsection{Principles, theories, models and heuristics}
\label{sec:principles}
What is Active Inference? One reading of it is as simply a principle -- minimise long-term surprise via a computable quantity called the \textit{Free Energy}\footnote{The \textit{Free Energy Principle (FEP)} \cite{friston2006free} suggests that any self-organizing system that is at equilibrium with its environment must minimize free energy, and is a mathematical formulation of how adaptive systems resist a natural tendency to disorder. It is composed of two terms $\mathbb{E}_{x\sim Q(x)} \ln P(y,x) - H(Q(x))$. The first is the energy, which is the surprise or information about the joint occurrence of the sensory input $y$ and its causes $x$. The second term is the negative entropy of the density.} that is hypothesised to explain the behaviour of sentient organisms. It could also be characterised as a theory to frame and represent problems of goal-directed behaviour (in the same sense as ``information theory'' or ``Bayesian theory''). This theory unifies perception and action by considering agents as if they stood at the end of predicted paths into all possible futures and looked back to infer which actions would be most likely to have taken along their preferred paths. It is not in itself a model, but leads to specific instantiations of models that can be implemented and tested. Many component parts need to be specified to instantiate a concrete model (Section \ref{sec:ainf_hci}). Active Inference leads to the heuristic of ``minimising surprise'' as a way to approach building intelligent systems.\footnote{It acts to \textit{minimise future surprise}, where surprise is defined idiosyncratically to have the pejorative sense of being both unexpected \emph{and} undesirable. In this framing, surprise implies being out of equilibrium; a human is ``surprised'' to find their body temperature to be 35\textdegree C and is motivated to avoid being surprised in this way.} This is a powerful but subtle way of thinking about interaction, which we develop further in §\ref{ref:min_surprise}. Active Inference does not specify or propose any specific algorithm but specifies how to combine probabilistic reasoning components to build agents. In that sense, it can be viewed as a framework for tying together particular machine learning algorithms into models that can be applied to interaction problems.

\begin{figure}
\vspace{-5pt}
\centerline{\includegraphics[width=0.95\linewidth]{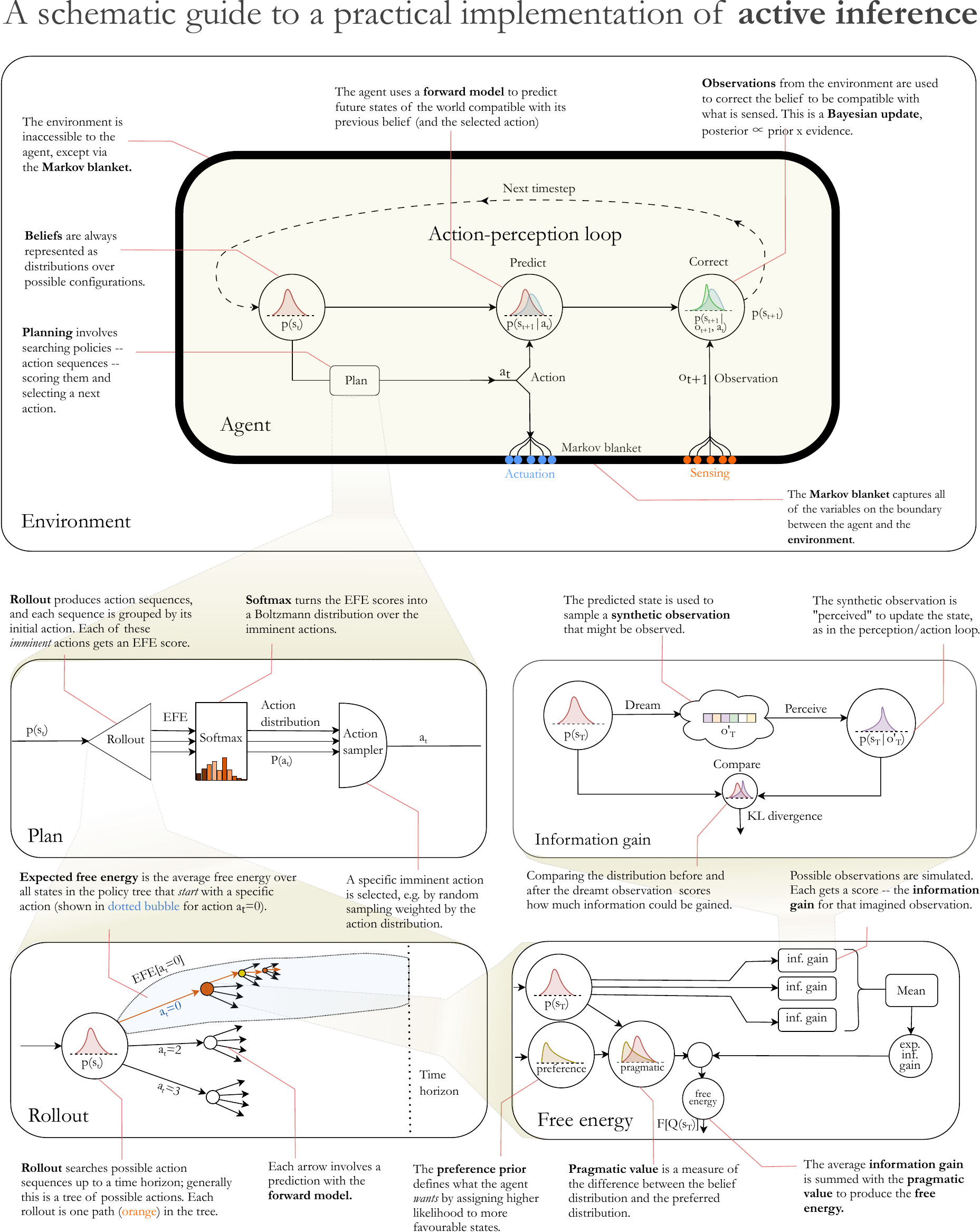}}
\caption{A schematic diagram of the Active Inference algorithm. An agents plans via a \textbf{rollout} process, scoring the tree of possible future states according to their information gain ("informativeness") and pragmatic value ("goodness") (See §\ref{ref:min_surprise} and App.\ref{sec:tutorial} for details). This is summarised into a single quantity for each imminent action, the \textit{Expected Free Energy (EFE)} and an action is sampled to minimise this value. The agent then updates its belief states using a Bayesian update given the sampled action and its perceptions from sensing.
}\label{fig:ainf_schematic}
\Description{Combination of Figures 2-6 in a single figure, showing the complete Active Inference diagram and how they fit together. }
\end{figure}

\subsection{Active Inference Agents}
\label{sec:AIF_agents}
We consider an agent 
-- an agent being a  entity distinct from its environment that perceives and acts with purpose -- embedded within a wider environment  (Fig. \ref{fig:ainf_schematic}, top). An Active Inference agent (AIF agent) is an agent whose actions are driven by the principles of Active Inference. Active Inference has three key aspects: 
\begin{itemize}
\item \textbf{Probabilistic} Active Inference is wholly probabilistic, and conceives of agents as entities that perform Bayesian inference to update their model of the world \textit{and} to decide upon actions. Both beliefs and preferences are encoded as distributions, not points.
\item \textbf{Predictive} Active Inference presumes agents who act based on predictions rather than on immediate perceptions. Predictive models decouple behaviour from immediate reactions to stimuli. Active Inference agents predict future beliefs over the environment and use their generative models to predict the sensations those beliefs would be expected to evoke.
\item \textbf{Unified} Active Inference combines reasoning about perception with reasoning about actions in a single unified inference problem. \citet{millidge2021whence} summarise it thus: \textit{``Intuitively, [Active Inference] turns the action selection problem on its head. Instead of saying: I have some goal, what do I have to do to achieve it? The Active Inference agent asks: Given that my goals were achieved, what would have been the most probable actions that I took?''}
\end{itemize}
The agent is explicitly isolated from its environment, except via a defined set of mediating states, the \textbf{Markov blanket} (Fig \ref{fig:markov}). 
This blanket is demarcated into states which actuate (modify the environment) and sense (respond to the environment). The behaviour of the agent is governed by three elements:
\begin{itemize}
    \item a \emph{preference prior} that defines which states the agent would prefer to be in. This means that its goals or desires are defined as a distribution rather than a single reference.
	\item a \emph{forward model}\footnote{
The use of a forward model in Active Inference is akin to a `digital twin' which allows the agent to monitor activity and infer hidden states (such as possible user intentions or goals), and predict possible outcomes to adapt the interface, make decisions, and act on the world.} that predicts how the environment evolves over time, conditioned on actions that the agent might take;
	\item an \emph{observation model} that predicts what might be observed (by the agent's sensors) in a particular state. Note that it does not react directly to sensation it perceives, but to its belief about the future states of the world.  The only role of observations is to correct estimates of beliefs about the environment.
\end{itemize}

The agent acts by predicting future beliefs, scoring each immediate action by the average free energy of \textit{all} beliefs consequent to that imminent action (the \emph{Expected Free Energy}, EFE) and then selecting an action to minimise the EFE, and thus expected future surprise.

\subsection{Active Inference in HCI}
\label{sec:AIF_for_HCI}
Active Inference is a computational approach that proposes a mechanism that drives intelligent behaviour. In the context of human--computer interaction, it is a computational interaction \cite{oulasvirta2018computational} approach, an approach focused on building and applying \textit{implementable models} in interactive systems. Like all computational interaction approaches, an AIF approach to HCI emphasises explicit and precise description of the interaction in a form that can be analysed \textit{in silico}. As Active Inference is not familiar to most HCI researchers, we provide a one-page visual summary of the core computational elements of AIF is presented in Figure \ref{fig:ainf_schematic}. We explicitly conceive of Active Inference from an algorithmic perspective as we intend these ideas to be implemented as practical software models. The details of the components in Figure \ref{fig:ainf_schematic} and an explanation of the algorithm are addressed in Appendix \ref{sec:tutorial}. The mathematical exposition is delegated to Appendix \ref{sec:maths}. 

One of the appealing aspects of Active Inference is that it unifies many strands of thinking in computational interaction. Active Inference is fundamentally a Bayesian approach, which conceives of reasoning as probabilistic updates, providing a guide for how rational agents should solve problems of induction, integrating prior knowledge with new data. This is well aligned with developments over recent decades in the cognitive science community, working on reverse-engineering the mind from Bayesian first principles \cite{griffiths2024bayesian}.

With respect to HCI, the probabilistic nature of Active Inference makes it well suited to deal with the uncertainty rife in interaction problems. Users and contexts of use are fundamentally uncertain; failure to reason about this uncertainty leads to fragility. The predictive nature of Active Inference is both useful when applied as a model of human cognitive processes (humans clearly exhibit predictive elements in their actions) and in also a systematic way to address latencies inherent in interactive closed-loops.




\subsection{Human--Computer Interaction}
\label{sec:ainf_hci}
Active Inference describes how agents may behave.  But in an HCI context, who is the agent? What is the environment?  

\subsubsection{Principles for applying Active Inference in interaction}

We now adapt the standard AIF model to represent humans and computer systems from an Active Inference perspective:

{\bf AIF Human model:} Our model of the human agent can predict what their senses will observe following actions they can make. The human agent can mentally make future plans and, using the expected free energy metric, pick actions which will reduce the surprise emanating from the computer, to bring the unobservable internal states of the computer to preferred values. Users observe computer displays as part of their observation vector. 

{\bf AIF Computer model:} In turn, human actions are observed by the sensors of the computer. The computer agent can generate its own candidate action plans which parameterise the computer's display to the human (e.g. change the display, make a sound, send a vibration), or actions on the shared environment. This could include manipulating robot limbs or environmental adjustments (e.g. lighting, heating or air conditioning). The computer's display actions feed into human agent's observations with a fidelity which depends on the quality of the computer's display, the human's perceptual systems and attention, and any disturbances from the environment. The AIF needs probabilistic forward models of sensors, interface dynamics and displays (video, audio, tactile).

{\bf AIF Interface model (transduction):} a variation on the general Active Inference model of the computer is to model an interface that looks in both directions to black box user and system models. This may be appropriate when the states of the computer are not accessible or reliably predictable for the interface elements, forcing it to view the rest of the system as an uncertain environment.

\subsection{Virtues of AIF for HCI}
What is the point? Why would one wish to build systems that are driven by Active Inference principles or to model users as Active Inference agents?  When is it appropriate to give an interaction object a {\it purpose}, and how will human users respond to that? Designing interaction objects as agents with purpose or goals has the potential to make them more resilient to changes in the context and, as William \citet{james1890principles} said, enable them to achieve stable goals via flexible means,\footnote{{\it ``Alter the preexisting conditions, and with inorganic materials you bring forth each time a different apparent end. But with intelligent agents, altering the conditions changes the activity displayed, but not the end reached; for here the idea of the yet unrealised end cooperates with the conditions to determine what the activities shall be.''} \cite{james1890principles}. } and can reduce the complexity that the user needs to control directly.

There are several virtues that typify Active Inference. Active Inference proposes a principle for action that is \textbf{robust}, in the sense that Active Inference agents apply Bayesian reasoning with appropriately quantified and propagated uncertainty. Much unmodelled complexity can be absorbed into uncertainty, as opposed to building increasingly intricate but fragile models. Active Inference emphasises \textbf{resilience}, specifically the framing of goal-directed behaviour as the drive to remain in ``comfortable homeostasis''.  We would therefore expect AIF-powered interfaces to have superior qualities in remaining stable and controllable in a wide variety of conditions. As a Bayesian approach, Active Inference is \textbf{parsimonious} -- especially powerful in small data regimes, where sparse and noisy data can be efficiently combined with strong prior models. This is in sharp contrast to conventional deep learning methods that rely on extensive historical datasets for learning; this restricts conventional ML in interaction contexts where such data is hard to come by. It is particularly challenging for traditional ML to deal with closed-loop adaptation because observations of a particular user in a particular context will always be sparse. 
Active Inference models are also fundamentally \textbf{predictive}; they operate exclusively on predictions of possible futures. This is a useful feature when modeling human users, to emulate the predictive nature of human behaviour. It is also of interest in building interactive systems, where predictive aspects can account for latency and beyond that lead to anticipatory interfaces. An Active Inference based approach is also \textbf{adaptive} in that Active Inference involves reasoning ``in the moment'' -- updating models and their parameters not solely in an offline training regime but during the evolution of behaviour itself. This affords a class of intelligent behaviour that might be harder to realise with approaches that rely on amortised inference.

\subsection{Active Inference configurations in Interactive Systems}
\label{sec:configurations}

\begin{figure}[htb!]
\includegraphics[width=\textwidth]{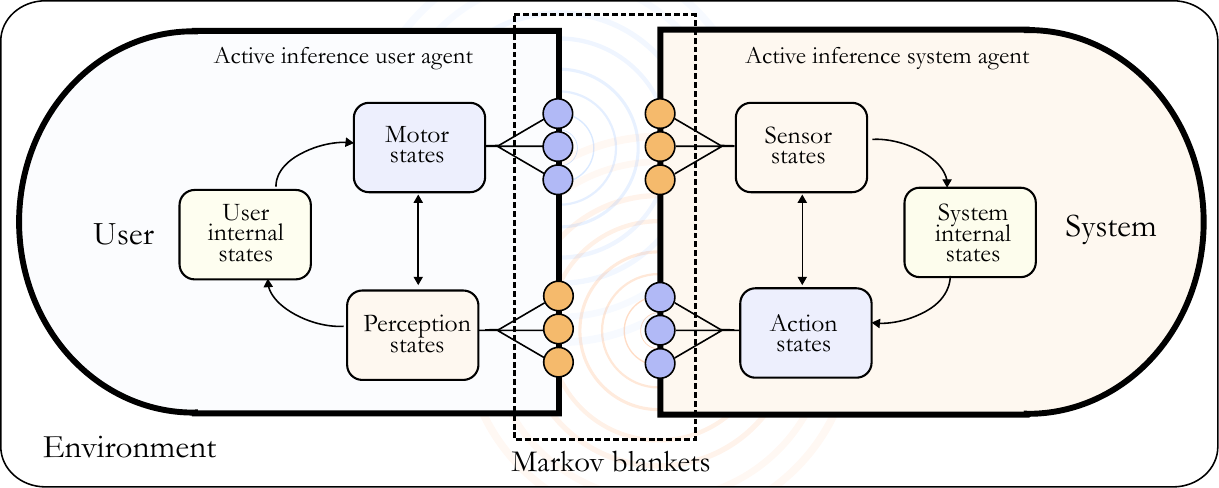}
    \caption{The Active Inference representation of the entire human--computer interaction loop as a dyad of mutually interacting agents. Each agent can only perceive or act on the environment via those variables that form its Markov blanket. Each agent is embedded in the environment of the other, and its actions impinge upon its partner's Markov blanket only via this environment.}
    \Description{The Active Inference loop with an Agent representing the User interacting with an Agent representing the System within an Environment}
    \label{fig:AIFloop}
\end{figure}

The {\it `Mutual Interaction'} offline simulation scenario, shown in Figure~\ref{fig:AIFloop}, takes an AIF approach to model the \textit{entire} interaction loop with a dyad of a human user and a computer system, both of which are modelled as AIF agents. 
Each agent has its own goals/preferences and can learn during their interaction. The user and system become each other's environment (we discuss the environment in more detail in §\ref{sec:environment}).
Both agents dynamically interact with each other to align their observations with preferences, but also engage in the necessary precursor of minimising uncertainty about hidden states in the partner agent. 

By modelling the whole loop, we introduce the benefits of simulation (§\ref{sec:simulation}) and open up the possibility of computational solutions to explain interaction (§\ref{sec:explan_power}), including  open challenges such as measures of {\it agency, engagement} (§\ref{sec:eval_power} and §\ref{sec:causality}), the {\it boundaries in Human--Computer interaction} (§\ref{sec:markov}), and how best to represent human {\it preferences} and {\it goals} (§\ref{sec:preferences}).

\begin{figure}
    \centering
    \includegraphics[width=\linewidth]{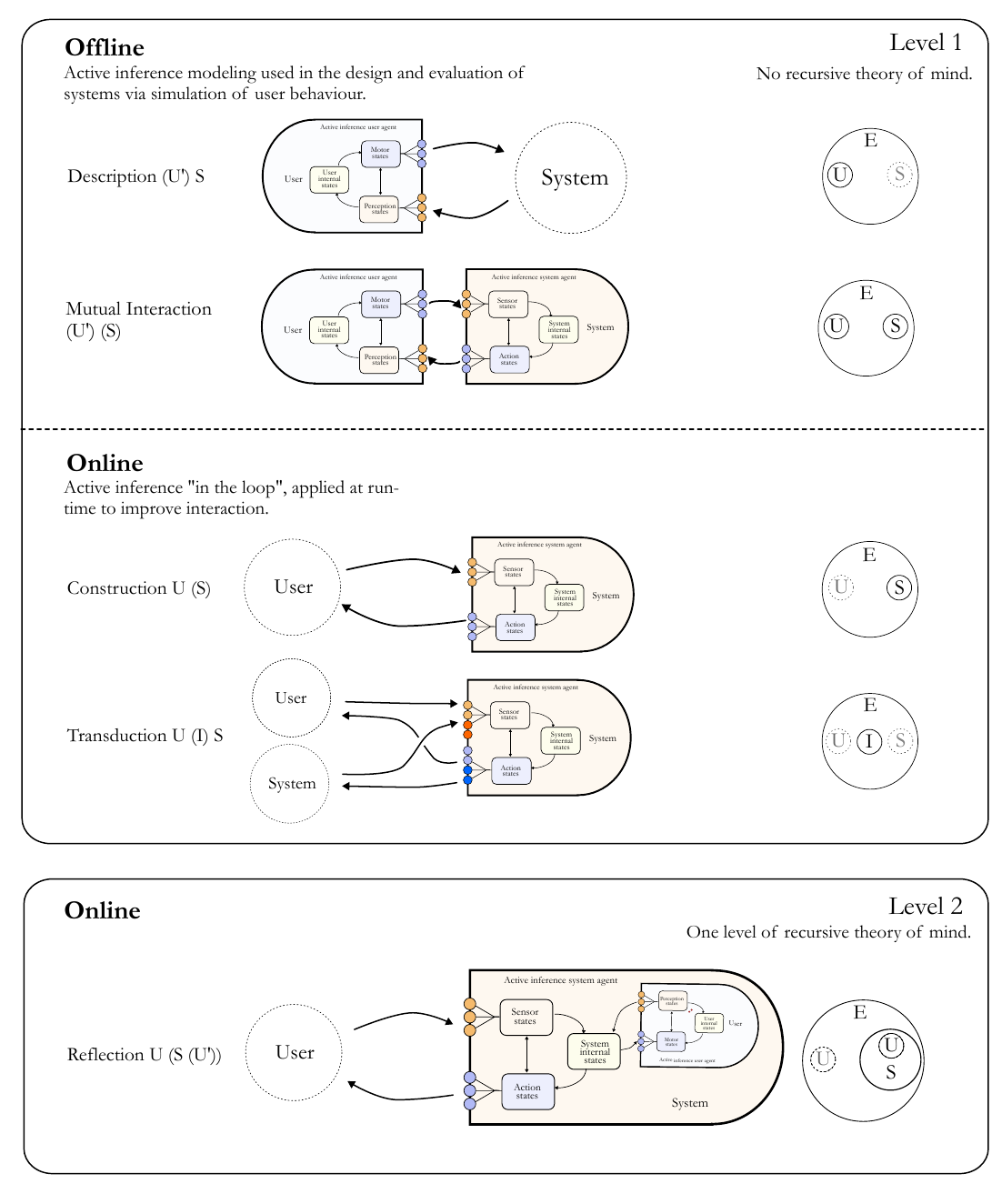}
    \caption{Different application modes of Active Inference. Active Inference can be used to simulate user behaviour, or joint user-system behaviour (``offline''). Alternatively, systems can be built that apply AIF in the interaction loop (``online''). Increasing levels of sophistication involve additional nested active inference agents, as in the reflective agent that operates using an internal AIF simulator of a user, as shown in the lower box. \texttt{E} indicates environment. Dotted circles are non-AIF units.}
    \label{fig:online_offline}
\end{figure}

We can, however, apply ideas from active inference to HCI in other  configurations where non AIF systems or real users are included in the loop, as shown in Figure~\ref{fig:online_offline}, and described below.\footnote{Our textual descriptions of these configurations use parentheses to indicate the part of the user-system loop modelled as an Active Inference agent. Primes (like U') indicate ``model of'' (for example, \texttt{U'} means ``model of the user'').}

\subsubsection{Offline simulation and analysis of interactive systems via AIF}
There are configurations intended for offline simulation and analysis, where the user is modelled by an Active Inference model:
\begin{itemize}
	\item \texttt{(U') S} \textbf{Description} we can model a user \texttt{U} as operating under AIF principles to reason about how users might behave. The system \texttt{S} is part of the environment. This formalism can be used at design-time to simulate user behaviour and thus optimise designs, or at evaluation-time to interpret observed interaction behaviours.
    \item \texttt{((U') S)} \textbf{Mutual interaction} We can model the whole interaction loop, where both the user and the system are represented as AIF agents and the joint behaviour of the mutually interacting agents is simulated.
\end{itemize}

\subsubsection{Online use of AIF in interactive systems}
\label{ref:online}
There are also configurations where the system or its interface are represented by an active inference model interacting with a real user. These are:
\begin{itemize}
	\item \texttt{U (S)} \textbf{Construction} we can build an interactive system \texttt{S} that acts by implementing Active Inference. Such a system would be a predictive, probabilistic interface. The user \texttt{U} is then a part of the environment but not explicitly modeled.
	\item \texttt{U (I) S} \textbf{Transduction} We can construct an active inference agent \texttt{I} who lies between a user and an existing system and mediates their interaction (neither of whom are assumed to be AIF agents). The user and the system jointly form the environment of the mediating agent. Our earlier work \cite{SteWilMur24} provides an example of such a model.
\end{itemize}

\subsubsection{Reflective Active Inference}
All of the active inference models above can be augmented by explicitly incorporating an active inference model of their partner within their generative forward model. 
\begin{itemize}
	\item \texttt{U (S (U'))} \textbf{Reflection} We can construct an AIF interactive system S that incorporates an AIF model of a user \texttt{U}.\footnote{The basic Construction model above changes from \texttt{U (S)}  to \texttt{U (S(U'))} when the system \texttt{S} has an explicit AIF model of the user. The mutual interaction case becomes \texttt{(U' (S')) (S(U'))}, and if we added a further level of mutual prediction it would become \texttt{(U' (S'(U''))) (S (U'(S'')))}.} The interface plans actions informed by a forward model that includes a user model, that is in turn an AIF model.
 \end{itemize}
 
Such reflective interfaces are examples of \textbf{mutual models}. \label{sec:mutual} For the human to model what they will observe given a specific action, they need to model the computer's likely responses to their actions; which, in turn, are based on the computer's model of how the human will respond to \textit{its} actions. Similarly the computer's model of its path from action to observation must include a model of the user's likely response, which will also include the user's model of the computer.  Mutual modelling and recursive theories of mind in interactive systems are reviewed by \citet{keurulainen2024role}, who describe a series of levels of mutual modelling.\footnote{Summarising the presentation in \cite{keurulainen2024role}: A Level 1 agent is invariant to the internal state of the partner agent. An agent which adapts its interaction based on beliefs about the level 1 partner model is a Level 2 agent. A Level 3 agent acts conditional on second-order nested beliefs about its partner, e.g. incorporating a model of their partner's model of themselves. This would allow a human to infer the knowledge an AIF system has about them by the actions it makes. A Level 4 system modelling the user at Level 3 could use Theory of Mind to understand the human's responses, and can act in ways that make the agent's beliefs identifiable to the human. Level 5 users are able to infer the intention behind system actions.}  This is shown in the lower box of Figure~\ref{fig:online_offline}.

 {\it `Sophistication'} is used in the economics literature to refer to having beliefs about one's own or another's beliefs, and \citet{friston2021sophisticated} point out that {\it ``most current illustrations of Active Inference can be regarded as unsophisticated or naive, in the sense that they only consider beliefs about the consequences of action, as
opposed to the consequences of action for beliefs''}. The ability to reason about beliefs will be important in interactive systems because the system has uncertain beliefs about user goals and preferences, and a key goal of interaction is to reduce the uncertainty the system has about these beliefs, so that it can act to support the user, given its knowledge about their goals and preferences.

\section{Core elements of Active Inference in interaction}\label{sec:distinctive}


Section~\ref{sec:AIF_agents} introduced the key features of general AIF agents. This section explores how these general elements need to be adapted to apply AIF thinking to HCI design, construction and analysis tasks. We develop each point in more detail in the following linked subsections.

\begin{itemize}
    \item \textbf{Agents} As discussed in §\ref{sec:ainf_hci}, we must decide which entities will be modelled or implemented as AIF agents: users, systems, both, neither? 
    
    \item \textbf{Environment} We need to establish the role of the environment. Is it just a ``transmission medium''? Is the user applying the system as an instrument to act upon or better sense the environment? Are both agents engaged in fighting environmental flux cooperatively? We discuss this in §\ref{sec:environment}.
    
    \item \textbf{Markov blankets -- Separating Agents from their Environment} Identify the actions and sensing each agent is capable of, to be able to define the boundaries between the user and system, the `Markov blanket'. §\ref{sec:markov}. 
    
    \item \textbf{Forward model} Construct a generative forward model. This is generally the most challenging implementation step, and involves both building a model that can predict future states \textit{and} synthesise the observations that would occur under those conditions.  In an interaction setting this includes models of human action, perception and cognition, as well as the system's sensing, computation and action. §\ref{sec:forward}

    \item \textbf{Prediction and reasoning about the best action} §\ref{sec:prediction} Active Inference relies on these forward, predictive models to estimate \textit{future} states, and predicates actions on those future estimates, driven by a intrinsic goal to reduce prediction error and \textbf{minimise surprise}. We  need to frame the problem in terms of surprise minimisation §\ref{ref:min_surprise}. This is the most challenging \textit{conceptual} step -- surprise minimisation is a powerful tool for thought, but it is sometimes counter-intuitive to apply.
    Computationally, each agent generates candidate policies for action sequences over their prediction horizon and selects one that minimises the Expected Free Energy. 
    
    \item \textbf {Preference priors} §\ref{sec:preferences}. An agent's goals or desires are defined as a distribution -- \emph{a preference prior} -- rather than a single reference. We therefore must define a preference prior to shape the actions of the system's agent to align with the goal of the interaction design. While surprise minimisation provides a degree of intrinsic motivation, aligning that with the intended purpose of the system requires explicit modelling of preferences.  This can be seen as a process of constructing attractors to drive the agent into preferable states.

    
    \item \textbf{Implement} This will involve building/training the forward model, implementing a rollout policy evaluator, and an inference mechanism to perform the Bayesian observation updates. We discuss the challenges involved in this in §\ref{sec:challenges}.
\end{itemize}







\subsection{The role of the environment}
\label{sec:environment}


\begin{figure}[htb!]
    \centering
    \includegraphics[width=\linewidth]{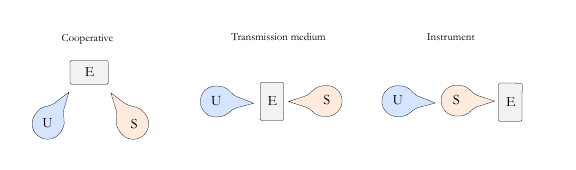}
    \caption{The environment in an AIF--HCI loop can be a) Something which is jointly controlled by a human and AIF system. b) A transmission medium c) Something which the user  observes or controls via the AIF system }
    \label{fig:environments}
\end{figure}

Our approach in §\ref{sec:configurations} differed from traditional presentations of Active Inference which describe a single agent in an environment. In our mutual interaction models, the computer and human AIF agents become the environments for each other. However, we may still need to explicitly model the wider environment beyond the simple human--computer dyad.

\subsubsection{Transmission}
The simplest extension is to include the {\it transmission} aspects of the environment between the system and user. This would involve forward models of the environmental impact from user action to system sensor and from system display to user perception.

\subsubsection{Intermediary instruments/actuators}
Is the user applying the system as an intermediary {\it instrument} to act upon or better sense the environment? Systems acting as \textit{intermediary instruments} or \textit{actuators} can support a user by transducing user-unobservable information from the environment into human-compatible percepts (as a microscope does), or it can act upon the environment at the behest of the user (as a robot arm does).\footnote{\textit{Inverse instruments} can also be imagined, where the human acts as a sensory organ for the system, relaying unobservable features of the environment, or translating digital displays into physical actions that the computer is unable to actuate.}

\subsubsection{Direct engagement with the environment}
Can both agents engage with the environment directly in fighting environmental flux cooperatively? For example, a smart speaker might be operated by a user to change the mood and observable behaviour of a party, but the user can also interact directly with others in the party. A central heating system application might be for controlling the temperature in a building, but the user can make other changes such as opening or closing windows.

In some interactions with the environment, agents will act to change the environment in future-oriented ways which maximise Expected Free Energy. This can be viewed as an example of {\it niche construction}, which is commonly observed in biological agents \cite{odling2024niche, bruineberg2018free}, where we can put effort into current actions which will simplify our future environment and make our existence more predictable and controllable. Such environmental action and perception also permits \textit{stigmergy} \cite{theraulaz1999}, where the environment is configured as augmentation to cognition or as an indirect mode of communication between the agents, as it can stimulate the same agent or its partner(s) to behave in a particular manner. HCI examples include XR applications, or even simple clustering of documents on a desktop.

\subsubsection{Multi-agent environments}
\label{sec:multiagentenv}
The environment models discussed so far have not had goals or preferences. However, in some cases the environment could include other humans or active, intelligent agents, which could be explicitly modelled by expanding the dyadic relationships in §\ref{sec:ainf_hci} to multi-agent relationships, or the complexity could be just kept as an undifferentiated part of the general environment forward model without representing individual agents.

 \subsection{Markov blankets -- separating agents from their environments}
 \label{sec:markov}
 Finding meaningful boundaries between an agent and its environment is a core aspect of Active Inference, and is a fundamental and controversial question of general interest for HCI, especially when AI assistants can take agency from humans for certain tasks. We will discuss what Markov blankets are, how they change, and why that is of conceptual and practical interest for HCI.

 \subsubsection{Markov Blankets}
\begin{figure}[htb!]
     \centering
     \includegraphics[width=\linewidth]{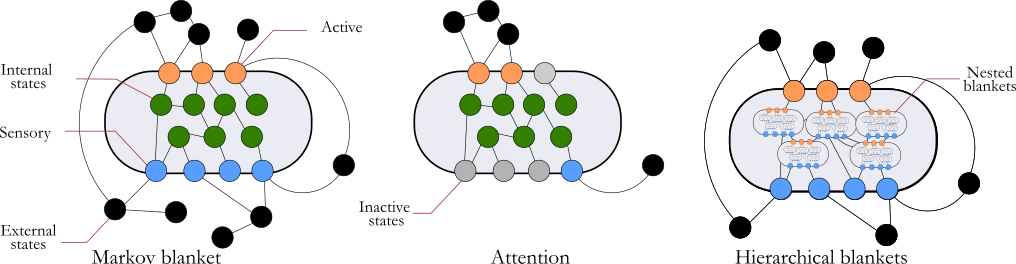}
     \caption{A Markov blanket statistically separates internal from external states; information flows exclusively via the active and sensory states. Attention can be seen as selective weighting/disabling of blanket states (center). Complex agents may have hierarchical Markov blankets (right).}
     \label{fig:markov}
     \Description{Three figures of Markov blankets, showing internal, external, sensory and action states. Attention figure has some greyed out states which are ignored. Hierarchical figure has sub-Markov Blankets within the Markov blanket instead of internal states.}
 \end{figure}

A {\it Markov blanket} defines the boundaries of a system in a statistical sense \cite{Pea88}. The existence of a Markov blanket means that there are external states that are conditionally independent of internal states and vice versa (Figure \ref{fig:markov}), and there are mature algorithms to support the inference of these states in Bayesian networks, e.g. \cite{aliferis2010local,pellet2008using}.  In AIF, a Markov blanket is the set of variables which mediate all statistical interactions between an agent and its environment.  Blanket states sit on the boundary, and within it we further distinguish {\it sensory} and {\it active} states. Internal states only affect external states via active states. Similarly, external states must pass through sensory states to affect internal states.  
This use of quantitative approaches to infer Markov blankets to represent the boundaries between an agent and its environment, could potentially be used to infer boundaries within the human--computer interaction loop.

\subsubsection{Where are the boundaries in human--computer interaction?}
\label{sec:boundaries}
\citet{hornbaek2017interaction} highlight that different approaches to interaction vary in how they portray the human--computer boundary, and how important that is for analysis and design, and we review that literature and discussion in §\ref{sec:boundary_related}

Technology can change the boundary. By giving a user a tool or assistive technology we transform their action and sensing capabilities. 
Modern tools will include context-sensitive algorithmic support from artificial intelligence, augmenting the user cognitively, or relieving them of the need to attend to, or act on mundane tasks.\footnote{Technologies such as deep neural networks and large language models are powerful ways for technical systems to interpret images, sound and text. This complex algorithmic manipulation of sensor observations has the potential to dramatically extend our perception, cognition and actuation capabilities, but it needs to be controllable by the user for successful interaction. Active Inference's standard mechanisms can model and predict such algorithmic transformations of the action--perception loop, allowing the agent to adjust its uncertainty in managing the complexity involved (e.g. how well does the ML-based classification of a file's music genre map to the current user's subjective assessment?).}
In cases where a user's behaviour indicates they wish to hand over some share of autonomy to an intelligent tool, it  effectively `pulls in' its Markov blanket, no longer controlling certain action states, or attending to (and having to predict the values of) certain sensory states.\footnote{We discuss how {\it shared autonomy} could arise in semi-autonomous driving in Appendix~\ref{sec:driving}. }

As discussed in \cite{HorOul17} the role of boundaries is a core and disputed aspect of competing theories of interaction, but they give no concrete way to objectively analyse such boundaries. The dynamic reconfiguration of boundaries by intelligent technologies can be associated with loss of agency, autonomy and freedom, motivating analytic tools which can infer boundaries, and map how they change in different designs and contexts.

\subsubsection{Dynamically changing boundaries and generalised tools}  The Markov blanket is not necessarily identified with the physical extent of the agent (the ``\textit{skin-bag}'' \cite{clark2003natural}). Consider a skilled craftsman with specialised tools, where does the boundary lie when they perform a task? \citet{kirchhoff2018markov} argue that for complex systems to survive they need to be able to hierarchically assemble Markov blankets of Markov blankets (Figure \ref{fig:markov}, right), and that these do not need to align with the physical boundaries of the organism, but can extend to include sensory or cognitive aids.  
\citet{kirchhoff2021determine} say that the {\it `Markov blanket formalism specifies a boundary for the mind that is negotiable and can move inwards and outwards over time. We show how the Markov blanket concept can ... 
include elements external to the individual's body'}. 
As agents learn and adapt over their lives, change their environments and, in the human case, create tools and technologies that change the boundaries themselves, \citet{clark2017knit} argues for \textit{``multiplicity, flexibility, and transformability of those boundaries''} \cite{clark2003natural,clark2008supersizing}, and proposes that {\it  ``considered as cognitive agents ...
the sets of tools, strategies, and devices (neural, bodily, and  bio-external) that constitute us as the mindful beings we are undergoes dramatic alteration.''} 
and that we are {\it ``nature's experts at knitting their own Markov blankets ... 
self-organizing processes that constantly re-invent themselves, repeatedly re-defining their own cognitive, bodily, and  sensory forms''} \cite{clark2017knit}.

In the previous section we discussed how shared autonomy could relieve people from mundane tasks. However, relieving people of certain tasks is also effectively reducing their freedom, as it relates to that task. 
\citet{satyanarayan2024intelligence} view intelligence as agency (which they define as `the ability to meaningfully act'), and design as delegating constrained agency. They illustrate dyads of `co-fused agents', with a dynamically shifting `line-of-control' which can give or take agency from the human or AI agent. They discuss the design challenge for drawing a line of control between human and artificial agents, and highlight that {\it `the form and balance of agency is highly dependent on cultural  attitudes, individual preferences, and the specific tasks that are being jointly performed'}.
In his attempt to compare the nature of human and machine intelligence, \citet{lawrence2024atomic} argues that it is important for us to understand what the `atomic' core of humanity is, the elements which cannot be replaced by AI. So we can see that getting to the core of interaction concepts is coupled to understanding how the effective boundaries between agents change during interaction. Mutual interaction models, implemented with AIF agents allow us to computationally analyse interaction loops in a variety of contexts and counterfactual conditions to gain insight into how boundaries change under different conditions, and what implications that might have for design.




\subsubsection{Attention}
\label{sec:attention}
In agents with limited capacity, attention is a key means to allocate processing, and has revolutionised deep learning \cite{vaswani2017attention}. We may be \textit{physically} able to sense a stimuli but not be attending to it. Attention focuses transient information flows specialised for the current task with changing Markov blankets associated with temporary conditional statistical independencies (Fig.~\ref{fig:markov}, centre). {\it `Attentional mechanisms may thus be seen as driving the formation and dissolution of a short-lived Markov partitioning within the neural economy itself, temporarily insulating some aspects of on-board processing from others according to the changing demands of task and context'} \cite{clark2017knit}. We therefore need to model  Markov blankets that can grow and shrink depending on the mode of interaction, and the consequent shifts of attention.

\subsubsection{Summary -- What Markov Blankets can do for HCI}
Markov blankets have clear value as sharp intellectual tools for segmenting functional roles in interactive systems. Markov blankets could be \textbf{detected}; we can use formal statistical measures to identify  blankets from data,\footnote{How much data is needed depends on the task complexity. This is potentially challenging for small groups of users, but more feasible once instrumented systems are deployed with end-users.} informing us how the boundary between human and computer shifts in different contexts, offering a computational instrument to quantify what happens to users of sensory and cognitive augmentative technology as their boundaries shift over time.
\footnote{We can also potentially \textbf{define} constraints on Markov blankets to support interaction. However, it is not yet clear how best to represent this in interaction engineering, and to what degree it can be used to constrain adaptation or help support calibration.}

\subsection{Forward and inverse models in interaction}\label{sec:forward}

Active Inference agents use {\it predictive probabilistic generative models} which can predict belief distributions over future states. Predictive models propagate beliefs through time; observation models grant agents the power to dream of the sensations they would expect in those states. Useful predictive models may be very impoverished simulacra of the true environment; it is in no sense necessary (nor likely) that an AIF agent have a completely accurate model of its environment. 

In the interaction case we need forward models of human perception (what would the human perceive for a given state of the world) and the computer's sensing (what would the sensors read for a given human action). We also model the human motor control system (e.g. how reliable muscle actions are, such as signal-dependent noise or fatigue effects), and the reliability of the computer displays (taking into account environmental disturbances, such as bright sunlight obscuring  a screen, or audio displays being drowned out by loud sounds). 

\begin{figure}[htb]
\includegraphics[width=\linewidth]{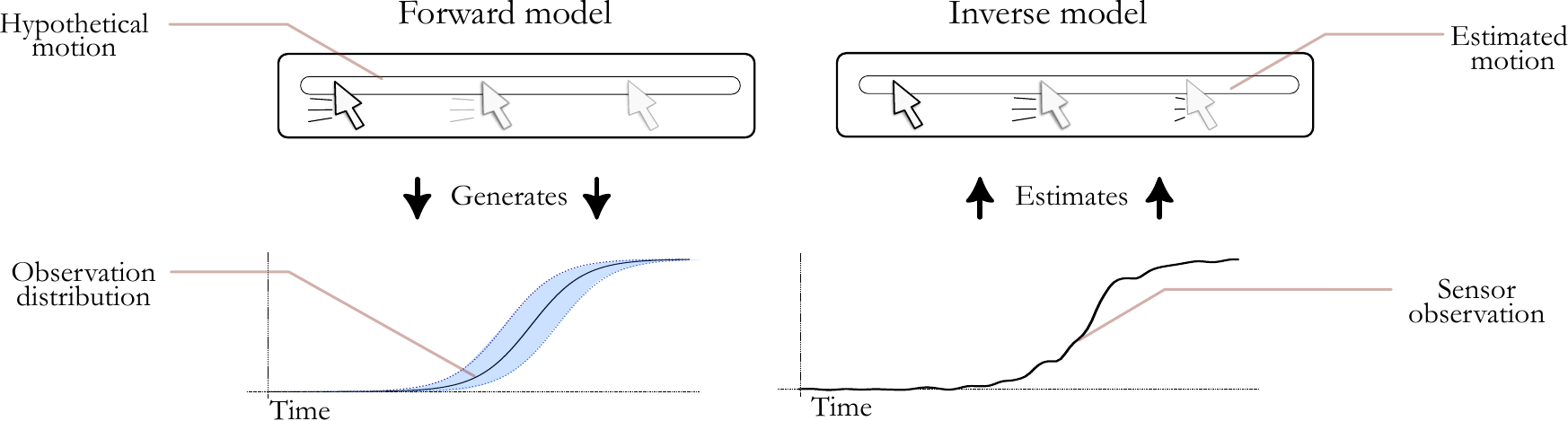}
\caption{Imagine detecting a cursor swipe gesture. Forward models (left) hypothesise what sensations (pointer trajectories, below) would be generated if a swipe were to be performed. Inverse models observe pointer trajectories and try to decode a specific swipe.}\label{fig:fwd_inv}
\Description{A visualisation of the forward and inverse models associated with a cursor swipe gesture, and their observed distributions.}
\end{figure}

\begin{figure}[htb!]
    \centering
    \includegraphics[width=0.9\linewidth]{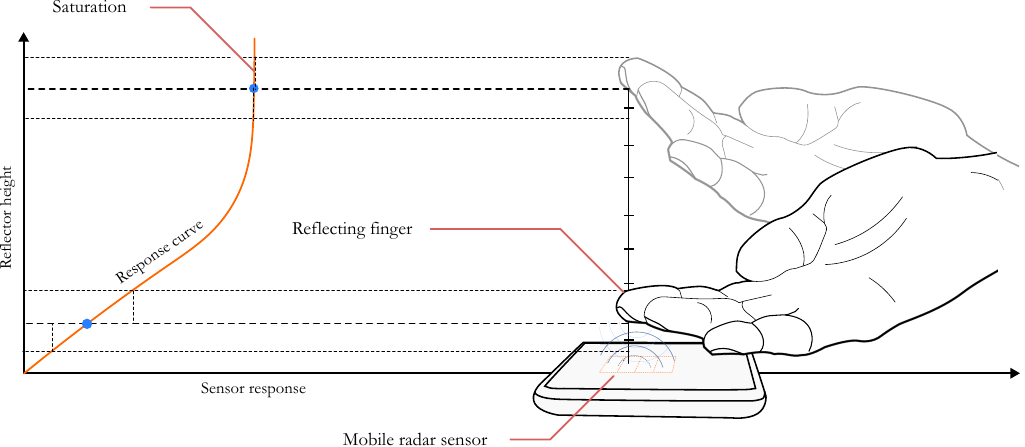}
    \caption{An example of ill-posedness in sensing. A mobile mm-wave radar detects the proximity of a hand. Close to the device, it is easy to resolve the true height of the hand from the radar signal. As the hand exceeds the sensitive range, inverting the signal becomes increasingly ill-posed.}
    \label{fig:hand_radar}
    \Description{A figure of a hand raised above a mobile phone. There is a response curve which flattens out when the hand reaches the limit of the radar's sensitivity.}
\end{figure}

Active Inference makes extensive use of forward models, so it is worth reviewing the contrast between forward and inverse modelling (Figure \ref{fig:fwd_inv}).  The {\it inverse problem} is the problem of using  observed measurements to infer latent parameters that characterise a system. This is a key problem in experimental science and engineering \cite{TikArs77,Tar05,Han10}. The {\it forward problem} is using models and their latent parameters to predict the observed measurements.
\citet{MurWilTon22} outline the role of forward and inverse modelling to support inference of human intention from sensors in HCI.   
Causal, forward models tend to be easier to specify, simulate\footnote{In many cases, where the scientific knowledge is mature, scientists can simulate a system better than they can effectively observe it.} and to gather experimental data for than inverse problems; but the inverse problem is what needs to be solved in an interaction context where we need to go from sensor readings to an estimate of the intention of a user. Bayesian approaches offer a general way to invert systems from forward models, using prior beliefs to constrain the solution set. \citet{MurWilTon22} point out that the inversion from sensing to intention may be robust to some variations in input, but very delicate in others 
because the inverse problem is ill-posed. For example, a nonlinear effect is present in the forward model of a mobile radar sensor (Figure \ref{fig:hand_radar}). As a hand moves away from the sensor we see a flattening response as the hand exceeds the range of the sensor. While we can predict the forward values with precision, there are infinitely many solutions to the inverse problem in  saturated areas. 


\subsubsection{Active Inference does not need inverse models}
Active Inference, in common with Bayesian approaches, sidesteps some of these invertibility issues with its generative approach which tracks latent states of the environment.
Active inference loops are exclusively built around forward models, and \citet{pio2016active} observe that a distinguishing feature is that active inference dispenses with inverse models: 
{\it "In active inference, there is no inverse model or cost function and the resulting trajectories are
Bayes optimal ... This contrasts with optimal control, which calls on the inverse model to finesse problems incurred by sensorimotor noise and delays. Inverse models are not required in active inference, because the robot’s generative (or forward) model is inverted during the inference. ... optimal control formulations start with a desired endpoint (consequence) and tried to reverse engineer the forces (causes) that produce the desired consequences. It is this construction that poses a difficult inverse problem with solutions that are not generally robust ... Active Inference finesses this problem by starting with the {\em causes} of movement, as opposed to the {\em consequences}."} 

Furthermore, the closed-loop nature of the active inference-based interaction means that if there are regions of the state space where the agent \textit{can} end up uncertain, the agent implicitly acts to avoid these states to reduce its future uncertainty and concomitant sensory surprise. However, there is no free lunch. The success of this inferential approach will depend on the quality of the  generative model and the computational cost of applying it. 

\subsubsection{Sensing for interaction}
The impact of the previous section is that when using Active Inference, we do not build models to map sensors onto unknown states; we build models to map unknown states onto sensors. An AIF agent combines top-down prior information about likely causes of sensations, with bottom-up sensory stimuli, rather than mere bottom-up transduction of sensory states. This is a constructive process where sensations test causal hypotheses about how they were generated. The agent chooses actions which will, over time, reduce the level of surprise at its sensory organs.

We anticipate that generative modelling will help to address future interaction challenges, where we will need to deal with richer, high-dimensional, uncertain sensors which are less optimised for human input. Existing HCI methods often lead to awkward and inefficient interactions, tightly bound to specific implementation details. This is unlikely to be resolved by minor tweaks to established interaction paradigms. Active inference implies fundamental rethinking of what interaction is and how it should unfold.


\subsection{The role of prediction and minimising surprise}
\label{sec:prediction}


An AIF agent does not react directly to sensation it perceives, but to its belief about the future states of the world.  The only role of observations is to correct estimates of beliefs about the environment.

\subsubsection{Minimise surprise}
\label{ref:min_surprise}

Active Inference relies on the forward, predictive models discussed in §\ref{sec:forward} that estimate \textit{future} states, and predicates actions on those future estimates, driven by a intrinsic goal to reduce prediction error. 
The agent avoids surprise by ascribing a value, the \emph{expected free energy (EFE)} to future states and then selecting an action to minimise the EFE, and thus expected future surprise. The EFE is an upper bound on surprise \cite{Fri2009b} that is computable. This can be decomposed into two components (see eq. (\ref{eq:expected_free_energy}) in Appendix \ref{sec:maths}): 1. the \emph{information gain} characterises how much more could be learned (the epistemic gain)  about the environment in a hypothetical future; how much information could leak in through the agent's Markov blanket, and 2.  the \emph{pragmatic value} quantifies the degree to which the hypothesised state of the environment agrees with the \emph{preference prior} -- a measure of how ``good'' it is from the agent's perspective.\footnote{There are other ways of decomposing free energy \cite{ParPezFri22}, p28-33, but this is the most insightful in this context.}  We can see these as a reflection of the classic exploration--exploitation trade-off, and their relative magnitudes determine whether an agent's behaviour is predominantly exploratory or exploitative.
Paradoxically, the goal of minimising surprise \textit{in the long run} leads to curiosity-like information gathering behaviours in the short term. AIF agents will naturally actively acquire information about their environment to mitigate the risk of future surprises. It is better to turn on the lights in a dark room, even if what  appears may be unpredictable, than to risk later surprise from that which lies unseen.

 Expected free energy summarises the value of all the paths through future consequences of an action in a single number that bounds this surprise. An AIF agent is therefore continually engaged in running counterfactual, `what if?' simulations, scoring them consistently via EFE. A compelling aspect of Active Inference is that action and perception are unified to cooperate in optimising this single objective, rather than being treated separately, which is an inelegance in other approaches.\footnote{It should be noted that while we have focused on EFE as the most common bound on surprise, there are free energies other than EFE that also bound surprise, such as the free energy of the expected future (FEEF). The specific choice of these quantities is discussed in detail in \citet{millidge2021whence}.}

An interesting observation is that actions based on EFE minimisation use predicted future outcomes as {\it causes} for the decisions and actions in the present.  This is relevant for 
 interpretation of human actions, as people's actions often only make sense when explained by their  anticipation of achieving some future state of the world. We relate this to traditional causality discussions in appendix~\ref{sec:causality}.

 \subsubsection{Coping with latency}
 Interactive systems have notable delays, particularly in human perception and motor execution. In systems with delays, we need to be able to predict to retain stability. The predictive element of Active Inference simplifies dealing with and reasoning about latency in interaction. An AIF agent's actions are driven purely by its predictions of what will happen, and not directly by what is happening now. \citet{Fri11WhatIsOptimal} notes, for example, that an Active Inference approach sidesteps some of the issues with modelling delay in human motor control, and compares the approach to  optimal control methods which also include forward models \cite{todorov2002optimal,todorov2004optimality}.
 
\subsubsection{Adaptation via curiosity supports diversity}
The goal of an interaction engineer is to enable user and system to communicate intention via interaction,  accommodating diversity in human performance and preferences.
An AIF agent uses predictive models to estimate the expected free energy, which imbues it with curiosity to test interaction options (whether sensor interpretations, interface parameters, autonomy level or preference settings). The user 
experiences this as a proactive system offering them personalised ways to interact. Such exploration makes a system more resilient to inter-user variability, context or varying preferences. This comes at a cost, as generative models must acquire more information to reduce their uncertainty about hidden states. Active Inference offers us a clear mathematical framework for analysis of these trade-offs. 

\subsubsection{Prediction Horizons}
Identifying appropriate prediction horizons helps prune the planning to make simulation computationally tractable, and to terminate the recursion of mutual theories of mind (this is associated with the levels of recursion for the mutual prediction discussed in §\ref{sec:mutual}). This will be an important practical parameter when designing interactive systems using AIF models.
 
 

\subsection{Preference priors}
\label{sec:preferences}
\citet{hornbaek2017interaction} highlight that {\it `most [concepts of interaction] say little about how intentions are formed or affected by interaction'}. In AIF, intentions are encoded via preference priors\footnote{Sometimes simply referred to as ``priors''. This overloaded use is confusing and we avoid it.} over observations that the agent prefers to sense or, alternatively, over states that the agent prefers to be in. The preference prior biases the agent's model of its environment, making it appear less surprising to transition into preferred states and more surprising to transition into others. AIF uses cross-entropy to evaluate the alignment of estimated beliefs with preferences.\footnote{As  \citet{Alexander1964} (p.22) notes, it is much easier to define \textit{misfit} than identify an objective norm: \textit{``We should find it almost impossible to characterize a house which fits its context. Yet it is the easiest thing in the world to name the specific kinds of misfit which prevent good fit.''}} This is a flexible but easily computed measure of agreement. Preference priors are not necessarily static and can change over time as an agent adopts different priorities. In a simple example, an AIF model of a pointing task might have a preference prior that changes to favour a new target once a target has been selected.


\subsubsection{Rewards are not preferences}
AIF researchers emphasise the difference between objective rewards in RL and surprise minimisation. Rewards are conceptually regarded as a property of the environment and are therefore assumed to be static in nature. A subtle but important distinction is that rewards are related to outcomes, while preferences are an attribute of an agent's beliefs.\footnote{\citet{pio2016active} discuss this in the robotics context: {\it "Active Inference
    dispenses with cost functions, as these are replaced by the robot’s (prior) beliefs ... replacing the cost function with prior beliefs means that minimizing cost corresponds to maximizing the marginal likelihood of a generative model"}} Preference priors are commonly found to be more intuitive to set, without giving name to an ideal. In standard AIF work focused on biological systems the preferences are often relatively easily-defined, reflecting observations consistent with states that the agent naturally thrives in. Defining preferences for human behaviour is less trivial, and we anticipate that as AIF--HCI develops there will be significant research into eliciting and calibrating preference priors. 

\subsubsection{Eliciting and learning preference priors}
Preference priors can be learned from data, either online or offline. For example \citet{ShiKimHwa22PriorPrefLearn} present an approach to learn a prior preference of an AIF agent from expert simulation. \citet{sajid2022active} generalise reward learning, in which preferences are themselves learned,
‘learning to prefer’, to Active Inference. 
The interaction with an agent's niche and preference learning is discussed in \cite{bruineberg2018free}. \citet{sajid2021exploration} present Pepper, an AIF approach to a reward-free preference learning mechanism which allows an agent to interact with the environment and  develop preferences that it acts to satisfy without an extrinsic reward signal. 

\subsubsection{Misaligned preferences}
The simple one-agent setting has a single set of prior preference distributions. In the AIF--HCI case, we can model prior preferences in both elements of the interacting dyad. These preferences should be aligned, to ensure that the computer is acting in the interests of the human, as the purpose of the joint human-computer system is normally to support the user in achieving their goals. 

However, it is not necessary that preference distributions be mirrored perfectly.
The timescale over which we evaluate performance justifies differential preference prior structures. For example, in educational contexts, learning often involves suffering in the short-term for longer-term payoff. Actions that benefit the user may become apparent after many episodes of use, but the instructor agent would need preferences that appropriately shape behaviour within a single episode of use. 
Preference may account for others or society generally, as in a courteous and safe autonomous driving system 
or a preference to avoid expending energy. Misaligned preference may also arise because one of the agents has lower-level capacities with which it must concern itself, such as a vehicle that needs to maintain stability.

\section{Theoretical contributions of AIF--HCI}\label{sec:benefits}
We now discuss how Active Inference can offer a general theoretical framework for HCI, structuring our discussion along dimensions used by \citet{HorKriOul24} to evaluate the contribution of a `theory of HCI'. 

\subsection{Predictive power -- inherent to AIF}
As a computational simulation of \textit{all} aspects of the interaction loop, AIF inherently supports prediction of observable and latent states over time, at a range of levels of abstraction and timescales, and can extrapolate to generate predictions beyond regions covered in experiments. Prediction in such a probabilistic approach also includes the prediction uncertainty, which can be used both on- and off-line to improve the resilience of the interaction loop to disturbances, or new contexts. 

\subsection{Explanatory power: Studying the interaction loop with model-based analytic tools}
\label{sec:explan_power}

AIF can, from some very basic assumptions, provide coherent explanations for behaviours or phenomena observed in HCI. A specific novel explanatory strength is formally analysing effective boundaries between humans and computers via Markov blankets and using the models to measure agency or freedom in different contexts. AIF agents adapt their behaviour when uncertain about their environment, and this can often be a key factor in interpreting and understanding the behaviour of an intelligent agent, whether human or artificial.  This is of particular value in modelling shared control scenarios, where humans interact with AI algorithmic support.

The closed-loop, predictive structure of the Active Inference action--perception loop frames the human--computer interaction problem through a lens distinct from that in most textbooks (which often have separate chapters on `Input' and `Output'), highlighting the irreducible interdependence of action and sensing. 
Section~\ref{sec:causality} argues that interaction cannot be fully understood either from users, technology or their environment in isolation. Fundamental aspects of interaction can only be quantitatively modelled by considering the \textit{whole} interaction loop.  

\subsection{Evaluation power}
\label{sec:eval_power}

Active Inference uses explicit computational generative models of each component of the loop. This allows us to take any state of the system and run counterfactual experiments to evaluate degrees of freedom available to the agent. This can support conceptual breakthroughs and provide quantitative measures for concepts which were previously challenging to represent, such as agency, engagement, interaction and freedom.\footnote{A model-based approach is arguably the \textit{only}  way to realise quantitative measures of concepts such as engagement or agency, because of the need to perform counterfactual simulations of the consequences of different action options.}

As we discuss in more detail in §\ref{sec:causality}, \textbf{agency} relates to the degree of control one has over their environment, but it is tricky to model computationally. \citet{friston2012k} argue that when cost functions (as used in optimal control) are replaced with the free energy principle this grants agents a \textit{``sense of agency''}. Because Active Inference agents form a distribution over the future actions they \textit{may} take, they are directly computing their freedom to act. Their reasoning naturally includes predictions of futures which are highly constrained and few actions are reasonable, and other futures where the choice of action is broadly free. This highlights that the `causes' for predictive AIF agents are often directly attributable to their modelled distribution on \textit{future} states. 
This approach has the potential to give us a more coherent theoretical basis for analysing {\bf interaction} itself -- recently \citet{hornbaek2017interaction} observed that {\it `the term {\em interaction} is field-defining yet easily confused'}.\footnote{It is instructive to consider how a theory of Active Inference in HCI would fit the classifications of theories of HCI provided by \citet{oulasvirta2022counterfactual}. Active Inference immediately fits their classes {\it `Interaction as Control'} and {\it`Interaction as Information transmission'}, but because of the consequences of the predictive agent-focused approach, it would also incorporate {\it Interaction as tool use, Interaction as Optimal Behaviour}, and {\it Interaction as Embodied Action}.} 
\citet{pattisapu2024free} mapped emotional {\it valence} to utility minus expected utility and {\it arousal} to the entropy of posterior beliefs. This, along with the measures of attention discussed in §\ref{sec:attention} can be used to support new measures of \textbf{engagement}, where we can track which aspects of the environment the user is sensitive to, and how that affects their control behaviour.

Interaction {\bf freedom} can manifest itself in two ways: a free interaction loop should lead to the joint system being able to reach many distinct preferred states (diverse ends). But a free interaction should also grant the user many ways to reach states which are compatible with the user's preferences (diverse means).  
Systems maximising freedom are likely to have common features, such as the freedom to use sensors or actuators in different, personalised ways, and the freedom to choose the level of autonomy in a system with algorithmic support.
AIF is based on minimising long-term surprise, but this does not force us into a narrow homeostasis; the principle in fact motivates novelty-seeking, curious or exploratory behaviour, visiting possible states of the environment where it is uncertain to reduce future surprise.  AIF therefore provides a formal, rigorous framework for comparing competing adaptive models to expand users' freedom of interaction  \cite{seth2016active,schwartenbeck2019computational}. 

\subsection{Guiding measurement}
Many traditional HCI experimental methods use summary data such as time-to-completion or error rates for comparison. In contrast, AIF  uses generative modelling of phenomena at every level of granularity and can thus use a wide range of empirical behavioural and sensor data to refine and calibrate its models. In addition, when using Active Inference on the system side of things, the process of minimising the Expected Free Energy includes the information gain term which automatically picks actions which balance exploitation with the highest epistemic gains to reduce model and state uncertainty. 

\subsection{Informing design}

Our task as interaction engineers is to fabricate the inference, sensors and display mechanisms such that the two agents (human and computer) can interact constructively to achieve a goal. Relying primarily on forward models, AIF concepts can help us reason {\it offline} in a data-efficient, model-based, computationally implementable fashion about the interaction loop, and can be used {\it online} for real-time inference, adaptation and exploration. Design is hard, involving many trade-offs. Forcing designers to decide everything in advance, in ignorance of the specific user, context and task is excessively constraining. This can be mitigated by leaving some adaptation to deployment where users and systems can negotiate how best to interact.
It is also becoming clearer that purely offline development reaches a limit because the historical data gathered with an earlier version of the system to calibrate the models no longer represent the reality once the loop is closed with the new interface. In recommender systems, for example, a gap opened between research and practice, due to the vulnerability of the traditional approach of testing on historical logs of user interactions \citep{DBLP:journals/tois/JadidinejadMO22,10.1145/3397271.3401230}. Recommender systems that perform well on historical data often rapidly deteriorate when they engage with real users. Simulations with user models can help mitigate this risk at design  \cite{ie2019recsim,wu2023goal}, but being able to actively adapt models once deployed will often be critical to achieve closed-loop success.



\begin{revise}

\end{revise}


\section{Current challenges and next steps}\label{sec:challenges}

\subsection{General Challenges}
Active Inference will face similar criticisms to computational simulation approaches to HCI. These include the cost and complexity of developing models, and the (claimed) inability of models to adequately represent the cognitive and perceptual complexity of humans, especially the sensitivity of behavior to details of context. Other critiques have included the perceived failure of models to capture the physical and social context of interaction \cite{murray2022simulation}. Both sides of this argument agree that understanding humans is core to HCI and the argument is fundamentally about the level of description involved. Sheridan emphasises 
the importance of {\it denotative} model descriptions \cite{sheridan2016modeling}, which minimise the variability of interpretation, so that the field can agree what it is talking about. {\it ``The process of modelling forces one to think hard about the slice of nature under consideration, to ask and answer the question of what are the essential features of the structure and function"}. Modelling in AIF terms requires precise thinking that throws into sharp relief questions that are often fuzzily-defined.  Building generative models that validate against user behavior is an acid test of whether we {\it really} understand an interactive system. 

Humans are complex, social, diverse and ever-changing.  The evolution of mobile, wearable and XR interaction means environments for computing systems have become rapidly more diverse, and thus less knowable. These are challenges to all simulation-based approaches.  However, AIF systems do not require a complete and perfect  representation of the external environment. The agent's form allows it to become a statistical model of its niche, embodying statistical regularities of its world 
\cite{kirchhoff2018markov,clark2017knit}. A pair of AIF agents will bring their joint system into a mutually predictable region of the state space, rather than having to accurately model a wide range of `outlier' behaviour.

Any adaptive intelligent system faces challenges in dealing with co-adaptation of users, and the need to account for recursive theories of mind \cite{keurulainen2024role}. AIF faces these challenges, but its capacity to model \textit{both} user and computer in an interaction loop means that it can approach these problems computationally.

\subsection{Active Inference Challenges}
There are other challenges specific to Active Inference itself:
\begin{itemize}
\item \textbf{Computation} 
AIF is computationally intensive, particularly in action rollouts, and engineering effort will be needed to balance planning with real-time requirements of online applications, for example via amortised algorithms \cite{Kingma2013}. Even in offline settings, computational developments like dynamic programming \cite{paul2024a} and deep learning approximations \cite{paul2024a} will be required to realise agents with long predictive horizons.

\item \textbf{Implementation} 
AIF is an elegant theoretical framework, but we do not yet have the software libraries and computational abstractions to build, reason about, and deploy interfaces. Current implementations of AIF like \texttt{RxInfer} \cite{bagaev2023} and \texttt{pymdp} \cite{heins2022} are nascent and not yet easily applied to interaction problems. 
We also lack developed workflows \cite{gelman2020bayesian,chandramouli2024workflow} to apply Active Inference to interaction. We need principled, computational human-centric engineering design to capture human behaviour and acquire the data to build preference priors and forward models.
\item \textbf{Preference modelling} AIF eschews rewards for the flexibility of preference priors. Casting interaction problems in terms of these novel approaches to preference is flexible but unfamiliar. This will require developments in how preference priors should be elicited \cite{ShiKimHwa22PriorPrefLearn} and validated, and how generalised preferences can be fused from individual preferences.
\item \textbf{Uncertainty and prediction} Representing uncertainty can make interfaces robust but users can struggle to comprehend uncertainty or understand systems that represent it (Section \ref{sec:uncertainty}). Prediction is a vital component of AIF and can mitigate latency in the interaction loop. But prediction can be unsettling for users. The predictive, pro-active nature may make interfaces endowed with AIF models appear to be in the `uncanny valley' \cite{mori1970bukimi}.
\end{itemize}

\subsection{Anticipated progress}
What do we expect the immediate impact of AIF on HCI to be? Online use may be one of the first areas to show results, particularly the {\it transduction} approaches introduced in \ref{sec:configurations} which mediates the interaction between users and systems. First simulation results for these have already been presented \cite{SteWilMur24}. Use of AIF in system {\it construction} is likely to first appear as part of a more general system, where the AIF approach is used to improve a specific aspect of interaction, while others are designed with classical approaches.  

Offline use of AIF will be relevant for systems development and basic research. {\it Descriptive} AIF models of user behaviour for standard interaction mechanisms are likely to be made available over the coming years and can be used to test candidate systems designs. 
Simulating {\it mutual interaction} is a more advanced topic, requiring the complexity of Active Inference to work for both agents in the dyad, and is likely to be initially more relevant for fundamental research into interaction and agency. Reflection with mutually predicting AIF models and higher levels of recursion in mutual mental models will push the conceptual and computational limits even further, but will provide a valuable framework for the future to develop concepts and tools about hows humans interact with ever more intelligent machines.

\section{Outlook}\label{sec:outlook}

\subsection{Active Inference in HCI}

We introduced \textit{Active Inference} to an HCI readership, and presented how this can be a coherent framework for the human--computer interaction loop, representing users and systems with probabilistic generative models driven by intrinsic goals of minimising surprise. It opens up connections between HCI and modern thinking in cognitive neuroscience, philosophy and theoretical biology. 
Active Inference in HCI  builds on model-based {\it computational interaction} approaches, but brings specific novelties in preference distributions, forward modelling, surprise minimisation, and associated analytic techniques such as Markov blankets. 
There are extensive challenges to be addressed to bring Active Inference into interaction: modelling of human cognitive and perceptual processes at a sufficient level; developing software tools to build, evaluate and debug AIF models; and computational techniques to implement AIF efficiently.

\subsection{Why now?}
Control-based approaches to HCI have been limited by the need to replicate human perception, especially visual perception. Recent breakthroughs in near human-level perception via machine learning have changed this dramatically \cite{Mnih2015}. 
However, \citet{lake2017building} observe {\it `truly human-like learning and thinking machines will have to reach beyond current engineering trends in both what they learn and how they learn it.'} and propose combining the strengths of recent machine learning advances with more structured Bayesian cognitive models. Cognitive science researchers have found growing evidence of the benefits of this approach to structuring the research agenda, \cite{griffiths2024bayesian}, and we anticipate it will also be relevant for HCI. Active Inference provides a coherent framework for integrating the benefits of machine learning into HCI.
 Computational advances have rendered the appealing theoretical properties of Active Inference manifestly implementable.  Theoretical developments in the fast-moving AIF community have developed the basic principle to a sophisticated and widely-applicable theory in a remarkably short period of time. 

\subsection{Why is this exciting?}
Active Inference has aroused extraordinary interest across diverse scientific fields. Partly, this is because it is an elegant and internally-consistent theoretical model that unifies many strands of thought. It is compelling and satisfying from a purely intellectual point-of-view. But it can also be expressed mathematically and directly implemented in code: it is actionable at every level.  Applying Active Inference in HCI stimulates new ways of thinking, and particularly once we step out of the envelope of the conventional interaction modes where user behaviour and design have sedimented. Active Inference offers promise in engineering interactions which are truly new; when exotic sensors are integrated; when new modes of interaction are brought online; when new ML algorithms are developed. 

We believe Active Inference will lead to resilient interactive systems that are far less fragile than we have come to accept, especially in unpredictable contexts. Interfaces built on AIF principles will exhibit rich, adaptive and robust behaviour with \textit{``flexible means to achieve stable goals''}. Active Inference is an online way of reasoning. AIF-powered systems do not bake in pre-canned responses or pre-learned behaviours; they reason and act in the moment and provide a level of dynamism that ML approaches to intelligent agents struggle with. As a tool for quantitatively modelling the entire interaction loop, its promise is unparalleled. An AIF account of the interaction loop gives precise and actionable routes to understand fundamental questions about interaction, agency, freedom, and engagement. In many ways, the advantages and challenges mirror those of researchers applying Bayesian methods in cognitive science \cite{griffiths2024bayesian}.

There is such a rich and deep seam of ideas that this paper can only lay out the first steps in the broad intellectual landscape that lies before us. From visualisation to UI evaluation, user modeling to critical design, social signal processing to augmented reality, these ideas offer a fresh and inspiring perspective for every research area in interaction.

\begin{acks}
The authors received funding from the {\it Designing Interaction Freedom via Active Inference (DIFAI)} ERC Advanced Grant (proposal 101097708, funded by the UK Horizon guarantee scheme as EPSRC project EP/Y029178/1). R.M-S. also received funding from EPSRC projects EP/T00097X/1, EP/R018634/1, and EP/T021020/1. We are also grateful for research gifts from Google and Aegean Airlines. The authors would also like to thank the following colleagues for valuable feedback on earlier drafts: Daniel Buschek, Karl Friston, Conor Heins, Kasper Hornbæk, Markus Klar, Antti Oulasvirta, Aini Putkonen.
\end{acks}

\bibliographystyle{ACM-Reference-Format}
\bibliography{ERCnobold,references,localrefs,Ainf_CHI_24,seb}

\pagebreak  

\appendix

\section{Related work}\label{sec:related}
We review the theory of Active Inference and contrast it with closely related interaction theories. Active Inference is an example of a closed-loop, feedback-based approach to modelling agents.  We therefore review cybernetic approaches. Perceptual Control Theory focusses on how actions change an agent's perceptions. Empowerment-based methods (Sec.~\ref{sec:empowerment}) use information theory to create probabilistic models of the channel describing the agent's action--perception loop. Closed-loop systems make it harder to determine causality (Sec.~\ref{sec:causality}). These challenges also affect users' sense of agency in interaction loops.

Active Inference approaches to HCI are a specific example of {\it Computational Interaction} (Sec.~\ref{sec:compint}), and lean heavily on Bayesian approaches to the interaction loop \cite{williamson2022introduction}. They are also closely related, but distinct from, the recent work in reinforcement learning (RL) in HCI (Section \ref{sec:rl}), as well as simulation-based approaches (Section \ref{sec:simulation}).


\subsection{Active Inference}
The core ideas of Active Inference were originally developed by Friston \cite{friston2006free} under the name \textit{``the free energy principle''}. Friston and his collaborators extensively published on applications of Active Inference in biology \cite{karl2012free, friston2013life}, motor control \cite{adams2013predictions, friston2011}, robotics \cite{pio2016active, dacosta2022} and neuroscience \cite{friston2014anatomy,bogacz2017tutorial}. Active Inference has also seen growing interest in consciousness research \cite{seth2021being} and philosophy of mind \cite{clark2015surfing,clark2023experience}. These theories of consciousness posit that the mind continuously constructs its own experience of reality and updates it via discrepancies with real sensory input.
The first textbook on the subject, Parr et al. \cite{ParPezFri22} (p57) makes the case that Active Inference is a unifying theory which encapsulates earlier frameworks from cybernetics, neuroscience and psychology. Specifically, they argue that Active Inference unites and extends three theoretical approaches:
 \begin{enumerate}
 \item Enactive theories of life and cognition, which emphasise the self-organisation of behaviour as agents interact with their environments.
 \item Cybernetic, or control-based approaches where behaviour is {\it teleological}, i.e. is seen as being driven by goals.
 \item Agents contain predictive models of their environment which guide their perception and action. These approaches are compatible with {\it active perception}  \cite{gregory1980perceptions} and top-down {\it predictive coding} \cite{rao1999predictive,hohwy2013predictive}
\end{enumerate}

In neuroscience, predictive coding/predictive processing theory postulates that the brain is constantly generating and updating a mental model of the environment \cite{hohwy2013predictive,millidge2021predictive}. The model predicts input signals from the senses that are then compared with the actual input signals from those senses. Predictive coding is member of a wider set of theories that follow the Bayesian brain hypothesis \cite{knill2004bayesian}. A recent overview of applications of Bayesian inference in cognitive science is given in \cite{griffiths2024bayesian}.
\citet{seth2014cybernetic} expands the predictive processing approach in his {\it Cybernetic Bayesian Brain} work, placing an emphasis on predictive modelling of internal physiological states and engaging with enactive and embodied cognitive science approaches. \citet{Buz2019brainInOut} advocates that the brain's fundamental function is to induce actions and predict the consequences of those actions to support its owner's survival. Examples of predictive perception in interaction settings include {\it phantom vibrations} from undergraduates' phones \cite{drouin2012phantom} or pagers for medical interns \cite{rothberg2010phantom,sauer2015phantom}.



\subsubsection{Active Inference in HCI}
Although it has not yet been widely adopted, the first explorations of Active Inference in HCI are being published. A recent example of its application in HCI is \cite{SteWilMur24}, which implemented an AIF approach for 1-of-$N$ selection, a fundamental building block of interactive systems, in the context of noisy sensors (relevant for Brain Computer Interfaces), and the authors formulated the
interface as an independent agent charged with facilitating the flow of information, acting as an active transducer\footnote{An example of the {\it Transduction} model in Section~\ref{sec:ainf_hci}.} able to reason about the environment and user characteristics to optimise this flow, and therefore used Active Inference to enable reliable selection to be made with unreliable sensors. 

Other early explorations are emerging in viewing trust as extended predictive control,  such that it could be applied to analyse trust between humans and technological artifacts \cite{schoeller2021trust}, and in multi-agent collective intelligence at multiple scales \cite{kaufmann2021active}. However, to date there have been no publications reviewing the broad potential for Active Inference in HCI.

\subsubsection{Boundaries in HCI}
\label{sec:boundary_related}
\citet{hornbaek2017interaction} highlight that different approaches to interaction vary in how they portray the human--computer boundary. \citet{hollnagel2017diminishing} highlights that the very name human--computer interaction assumes that there is a clear boundary between human and machine, and that {\it ``system boundaries are well-defined, internal and external interactions are similar, and that humans and machines are reactive ... these assumptions ... are not tenable for intractable systems with tight couplings to their environment''}. 
So, because interaction links human and computer, it makes it challenging to analyse them separately. \citet{beaudouin2004designing} argues {\it ``Designing interaction
 rather than interfaces means that our goal is to control the
 quality of the interaction between user and computer: user
 interfaces are the means, not the end.''}. This is taken up further in  \citet{taylor2015after} and,  as discussed in §\ref{sec:causality}, has been a concern in engineering since the 1960's, e.g. the {\it crossover effect} in flight control \cite{McRJex67}. \citet{bergstrom2025dira} argue that the user interface is currently a fuzzy concept, saying {\it ``We also think that it is useful to separate some aspects of the user and some aspects of the UI''} and propose a model composed of four elements: Devices, Interaction Techniques, Representations, and
Assemblies (DIRA).

\subsection{Cybernetics, closed loops, perceptual control theory, agency, and empowerment}
\subsubsection{Perceptual control theory}
One cybernetic approach to modelling agent behaviour related to active inference is {\it Perceptual Control Theory} (PCT). \citet{Pow73} argues that we control our `inputs', our perception, rather than our `outputs' or motor actions.  
Closed-loop interaction techniques inspired by PCT were used in HCI contexts by \citet{WilMur04bAI}, which were subsequently further developed by others \cite{fekete2009motion,velloso2017motion}. An explicitly probabilistic interpretation is given by \citet{velloso2021a}. \citet{ParPezFri22} observed {\it "While in both Active Inference and perceptual control theory it is a perceptual (and specifically a proprioceptive) prediction that controls action, the two theories differ in how control is operated. In Active Inference but not perceptual control theory, action control has anticipatory or feedforward aspects"}.
\subsubsection{Empowerment}
\label{sec:empowerment}
Another probabilistic, closed-loop approach to analysing interaction is {\it `empowerment'}, proposed by \citet{KlyPolNeh05a}, which combines control theory and information theory to measure how much information an agent can send from their actuators, through their environment to their perception (the channel capacity). An agent which cannot perceive the impact of its actions has low empowerment. Maximising empowerment is a behavioural heuristic for selecting actions. This was applied to interaction in \cite{trendafilov2017information,Trendafilov2013AI}.

\subsubsection{Causality and agency in closed-loops}
\label{sec:causality}
Human predictive abilities mean that human behaviour is deeply intertwined with the systems they control. This is not easy to compartmentalise in the style of a classical control system block diagram. For example, \citet{McRJex67}  demonstrated  in their {\it crossover model} research  that human pilots adapt their behaviour to shape the properties of the whole closed-loop, changing their control behaviour when the properties of the aircraft change. The {\it joint cognitive systems} approach used by \citet{hollnagel2005joint}  highlights the need to examine the joint human--system interaction, rather than only the individual blocks.\footnote{\citet{Hollnagel1999ModellingProcess} views the human as a model of the process they are controlling.} We will return to this when trying to define boundaries in interactive systems, in Section~\ref{sec:boundaries}.

The mutually predictive nature of interaction loops makes it challenging to reason about causality. \citet{bunge2017causality} sees interaction as {\it reciprocal causation}, as in the gravitational attraction of two bodies, and proposed approaches based on a special type of causal
relationship: {\it mutual determination}, rather than a traditional {\it causal determination} approach.  Interaction cannot be divided and attributed to a human or a system alone. This makes analysis of interactive systems more challenging, and also affects users' perceptions of agency. 


\citet{bennett2023does} and \citet{limerick2014experience} review  how the HCI literature understands {\it agency and autonomy}, because of the importance of these for informing design, giving users a sense of ownership and control and the relevance for the modern opportunities provided by artificial intelligence. \citet{satyanarayan2024intelligence}  reviews the role of the concept of agency in interaction with Generative AI systems.   \cite{coyle2012did,bergstrom2018really} explore users' perception of ownership of actions. \citet{cornelio2022sense} look at agency in emerging technologies. 
\citet{tajima2022whose,kasahara2019preemptive,kasahara2021preserving} and \citet{veillette2023temporal} present a series of experiments examining how timing of interventions such as electrical muscle stimulation affect the user's perception of agency. \citet{friston2012active} highlight how in active inference a sense of agency can be linked to a probabilistic representation of control which is independent of the actual actions emitted, but which allows the agent to evaluate the consequences of possible actions. As pointed out in section~\ref{ref:min_surprise}, by conditioning current actions on future predicted distributions, once the past history has been incorporated into the generative model and an initial state estimate, Active Inference has a causal structure which is based on {\it future} predicted states, rather than past events.

\subsection{Computational interaction}
\label{sec:compint}
{\it Computational interaction} uses algorithms and computationally implemented mathematical models to predict, explain and enhance interactive behaviour \cite{oulasvirta2018computational}. It includes formal representations of designs as well as predictive models of users, environment and technical systems. 

\subsubsection{Uncertainty and probabilistic interfaces}\label{sec:uncertainty}
Active Inference puts uncertainty at the heart of interaction, and uses probabilistic, Bayesian methods to manipulate and represent it. Schwarz and her collaborators investigated probabilistic interfaces based on Monte Carlo sampling in a series of papers \cite{schwarz2010a, schwarz2011a, schwarz2015}, showing that modelling uncertainty can improve robustness. \citet{buschek2017} created a framework for probabilistic touch interfaces using hidden Markov models to improve resilience in touch-based UIs. \citet{weir2014} shows how appropriate probabilistic modelling of uncertainty in interaction can improve interpretation of user intention. Greis's thesis presents a wide-ranging review of the role of uncertainty in interaction \cite{greis2017}, including the challenges in communicating uncertainty to users.

\subsubsection{Simulation in HCI}\label{sec:simulation}

Active Inference is predicated on simulation models. In HCI, pioneering work in simulations was driven by \citet{card1983psychology}, whose Model Human Processor divided the user aspect into cognitive, motor-behavioural and perceptual components \cite{card1986model}.
The increasing use of computational
models has led to a greater role of simulation \cite{murray2022simulation} in HCI.  The term {\it Simulation Intelligence}, as proposed in \cite{lavin2021simulation} involves the development and integration of the key algorithms necessary for a merger of scientific computing, scientific simulation, and artificial intelligence. The original paper focused on other areas of science and engineering, but their methods are highly relevant to the complex modelling challenges of HCI.  \citet{kristensson2021design} use modelling to replace extensive experimentation in optimising text entry system parameters.

{\it OpenSim} enables the modelling, simulating, controlling, and analysis of the neuromusculoskeletal system \cite{seth2011opensim}. 
 \citet{ikkala2022breathing} used a realistic biomechanical simulation of a human body, learning muscle-actuated control policies based on perceptual feedback in interaction tasks with physical input devices, which allowed modelling of more realistic interaction tasks with cognitively plausible visuomotor control.  This approach shares the model-driven approach to modelling the human motor system, perceptual system and the sensing system, but differs from AIF in the reward structure, inference mechanisms and learns a direct control policy rather than optimising over possible future states.

\citet{moon2024real} used simulated models which matched the accuracy of human 3D selections to speed up selection and reduce errors. Computational issues of human simulation models are addressed in \cite{moon2023amortized,moon2022speeding}.

\subsubsection{Utility maximisation and bounded rationality}

Related approaches involve utility maximisation \cite{payne2022adaptive}. \citet{oulasvirta2022computational} propose `Computational Rationality' as a theory of interaction. The core assumption is that users act in accordance with what is best for them, given the limits imposed by their cognitive architecture and their experience of the task environment. This theory can be expressed in computational models that explain and predict interaction. It offers a principled, model-based, basis for algorithms to drive both inference and planning in cooperative agents \cite{howes2023towards}.

\subsection{Reinforcement Learning}\label{sec:rl}
Reinforcement Learning (RL) \cite{sutton2018reinforcement} is concerned with learning agent behavior that maps directly from observable state to (a probability distribution over) next actions such that, in expectation, a cumulative reward (value or utility) is maximised. Value maximisation is performed using an estimated value function in model-free RL \cite{DBLP:journals/corr/SchulmanWDRK17}, or using a model of the environment in model-based RL \cite{deisenroth2011pilco}. RL relies on manually specifying an auxiliary exploration objective (e.g., $\epsilon$-greedy, entropy regularisation or intrinsic motivation) to trade-off exploration and exploitation over the course of training \cite{DBLP:journals/corr/abs-2109-00157}, which often requires careful fine-tuning in practice. In contrast, free energy minimisation in Active Inference provides a Bayes-optimal trade-off between exploration and exploitation \cite{sajid2021active}. In RL, behavior is shaped by specifying rewarding state-action pairs, which turns out to be hard to do correctly in practice, and goals may change over time \cite{carroll2021estimating}. Approaches for training RL agents without explicit reward function (Inverse RL) include behavior cloning, reward function learning \cite{ng2000algorithms,adams2022survey,ramachandran2007bayesian}, learning from human feedback \cite{casper2023open}, and maximising empowerment \cite{myers2024learning,du2020ave}. Mis-specified rewards lead to unintended, undesirable behavior \cite{skalse2022defining, turner2021optimal} and have motivated methods for reward refinement \cite{hadfield2016cooperative,hadfield2017inverse}. An active inference agent is encouraged to generate preferred outcomes in proportion to their probabilities, which is distinct from repeatedly producing the most valuable ones.

 
In RL, latent states are commonly treated as environment stochasticity instead of being inferred, as is done in Active Inference. In this case, the only mechanism for adapting behavior to unobservable changes is policy retraining. Model-free RL agents have to visit safety-critical states before it can learn to avoid them. By contrast, model-based RL and Active Inference can quantify the risk of visiting those states and plan to avoid them. The combination of Bayesian inference, Bayes-optimal exploration-exploitation trade-offs, and safe exploration makes Active Inference appealing for modelling interaction with humans. In principle, all RL algorithms can be recast as active inference algorithms with some aspects removed, \cite{friston2009reinforcement} and close parallels can be drawn to RL with intrinsic motivation \cite{biehl2018expanding,chentanez2004intrinsically} and reward learning \cite{ShiKimHwa22PriorPrefLearn}.


\section{An introduction to the Active Inference algorithm}\label{sec:tutorial}

\label{sec:algorithm}
\label{sec:basic_AIF}
 
\begin{figure}
\includegraphics[width=\linewidth]{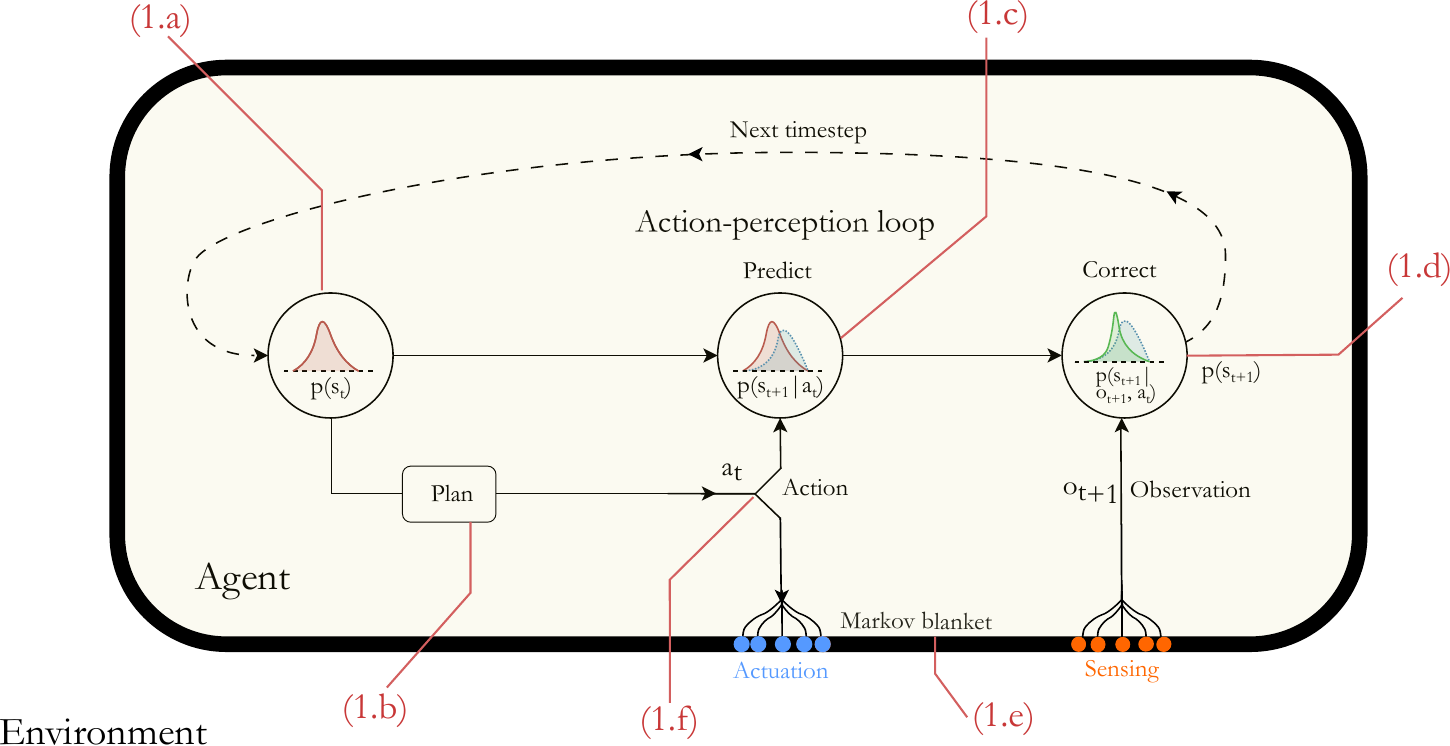}
\caption{An Active Inference agent, embedded within its environment. Beliefs are recursively updated using prediction and Bayesian updates, while actions are planned by rolling out hypothetical futures. The annotation labels refer to the discussion in Section \ref{sec:algorithm}. Figure \ref{fig:ainf_schematic} shows a diagram of how all of these components fit together.}\label{fig:ainf_agent}
\Description{A diagram of the Action-Perception loop of an Agent.}
\end{figure}

\subsection{Predict-correct}
The gross behaviour of an AIF agent is akin to a Bayesian filter (such as a Kalman filter) with an additional planning step that evaluates a tree of possible future actions and scores them according to expected free energy. The agent follows this pattern:

\begin{itemize}
	\item It begins with a distribution over possible states $p(s_t)$ (Fig. \ref{fig:ainf_agent} (1.a)) representing how it thinks the environment might be configured at time $t$.
	\item It evaluates possible futures to decide on an action $a_t$ (see \textbf{Plan} below).
	\item It executes this action $a_t$ on the environment (Fig. \ref{fig:ainf_agent} (1.f)).
	\item It uses its forward model to predict what the state of the world will be after this action. $p(s_{t+1}|a_t, s_t)$ (Fig. \ref{fig:ainf_agent} (1.c))
	\item It collects an observation from the environment, $o_{t+1}$.
	\item It uses its observation model to correct its estimate of the state of the world based on the actual observation  $p(s_{t+1}|a_t, s_t, o_{t+1})$, by performing a Bayesian belief update using its estimated state as the prior and the observation as the evidence to update on. (Fig. \ref{fig:ainf_agent} (1.d)). This update is classically performed with variational methods, though it does not have to be.
	\item This becomes the prior belief distribution $p(s_{t+1})$ for the next step of the reasoning process
\end{itemize}

Active Inference does not prescribe the specific algorithms to implement its component parts: forward models may be first principles algorithms, deep neural networks such as VAEs\cite{Kingma2013}, transformers\cite{vaswani2017attention} or any other learning algorithm; inference can be performed with variational, exact or Monte Carlo methods; rollouts can use fixed policy sequences, exhaustive search or Monte Carlo Tree Search \cite{coulom2007}. 

\subsection{Plan}

\begin{figure}
\centering
\includegraphics[width=0.7\linewidth]{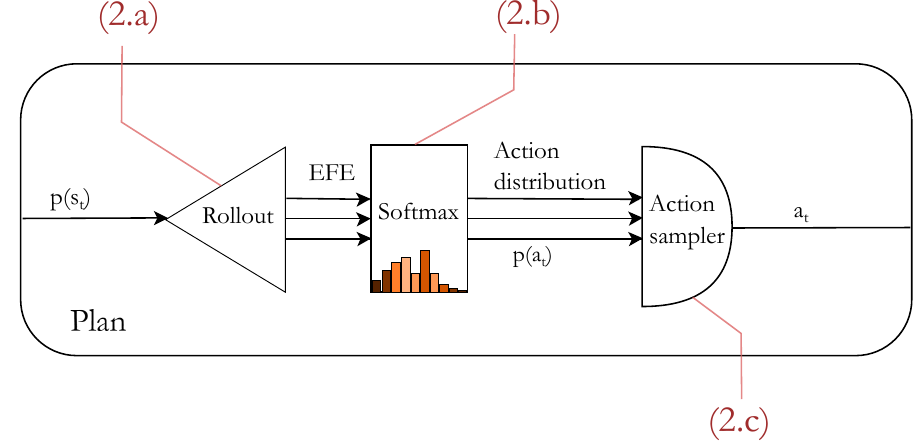}
\caption{The planning phase involves rolling out sequences of possible future actions, assigning them \textit{expected free energy} scores and then sampling actions based on these scores. }\label{fig:ainf_plan}
\Description{A diagram of the Planning phase with a Rollout, Softmax and Action Sampler}
\end{figure}

Planning (Fig. \ref{fig:ainf_agent} (1.b)) involves deciding upon an immediate action by estimating the consequences of all future states and potential actions consequent to it. In practice, the tree of future states is pruned to a limited set, for example using fixed time horizon.

\begin{itemize}
\item The \textbf{rollout} phase enumerates possible future action sequences (Fig. \ref{fig:ainf_plan} (2.a)).
\item Each imminent action is assigned an \textit{expected free energy (EFE)} (see \textbf{Rollout} below) 
\item These scores are negated and transformed into a probability distribution using softmax (Fig. \ref{fig:ainf_plan} (2.b)).
\item An imminent action is selected by drawing a sample from this distribution (Fig. \ref{fig:ainf_plan} (2.c)).
\end{itemize}

\subsection{Rollout}

\begin{figure}
\centering
\includegraphics[width=0.7\linewidth]{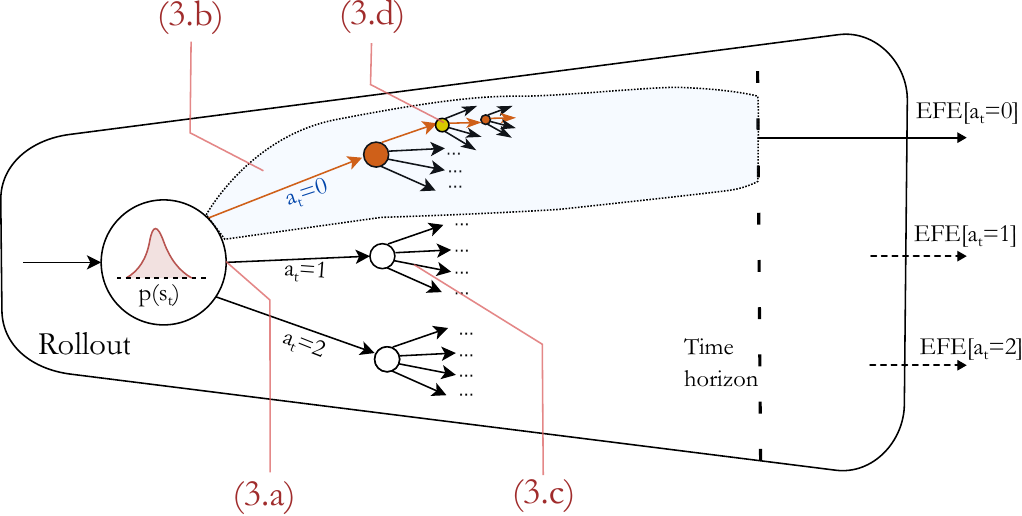}
\caption{Rollout traverses the tree of future actions. Each future state has a free energy computed. The expected free energy for an imminent action is the average of all free energies over all consequent states. }\label{fig:ainf_rollout}
\Description{A diagram of the rollout of a branching tree of future actions and the EFE associated with leaves of the tree.}
\end{figure}

During rollout, policies\footnote{Policies are simply action sequences in the Active Inference lexicon, rather than functions mapping states to actions, as in RL.} are evaluated and scored according to their expected free energy. 

\begin{itemize}
\item For each possible imminent action $a_t$  (or sample them, if the actions are unbounded). (Fig. \ref{fig:ainf_rollout} (3.a))

\item For $k$ until some time horizon $T$:
	\begin{itemize}
		
		\item For each action $a_{t+k}$
		\begin{itemize}
			\item \textbf{Predict} Using the forward model, estimate the distribution over hypothetical future states $p(s_{t+k}|a_{t+k})$. (Fig. \ref{fig:ainf_rollout} (3.c))
			\item A free energy is computed for each future state (Fig. \ref{fig:ainf_rollout} (3.b)). See \textbf{Free energy} below.
		  \item The search branches over the new set of available actions for the next timestep $a_{t+k+1}$.
	\end{itemize}
	\item Then we average the values over all of these futures, grouped by the \emph{immediate} next action $a_t$ -- giving the \emph{expected free energy (EFE)} for that action.  (Fig. \ref{fig:ainf_rollout} (3.b))
	\end{itemize}
 \end{itemize}

\subsection{Free energy}
\begin{figure}
\centering
\includegraphics[width=0.85\linewidth]{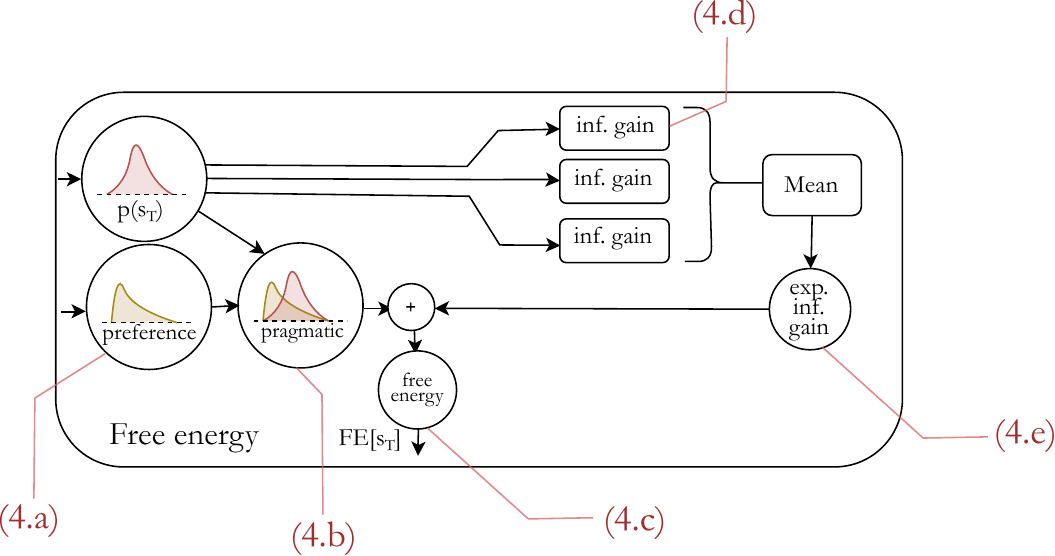}
\caption{The free energy of a predicted future state is a combination of how well the hypothesised distribution over states corresponds to the preferences of the agent (\textbf{pragmatic value}) and how much the agent expects to be able to learn from the environment in this state (\textbf{information gain}), estimated by sampling many synthetic observations.}\label{fig:ainf_free_energy}
\Description{A diagram of the Free energy calculation process}
\end{figure}

The free energy plays a fundamental role in Active Inference. It is a value that bounds the ``surprise'' of a belief distribution. It is computed by:
\begin{itemize}
    \item Comparing the predicted state distribution $p(s_t)$ to a preference prior $q(s_t)$ (Fig. \ref{fig:ainf_free_energy})  to get the \textbf{pragmatic value} of a state.\footnote{This comparison is the cross-entropy between the state distribution and the preference prior.} The preference prior is the key component that defines the behaviour of the agent. 
    \item Synthesizing observations that would be plausible under the predicted state distribution (Fig. \ref{fig:ainf_free_energy} (4.d))
    \item Computing the \textbf{information gain} for those hypothetical observations (see below)
    \item Averaging all of the information gains to estimate the expected information gain (Fig. \ref{fig:ainf_free_energy} (4.e))
    \item Summing the \textbf{pragmatic value} and the \textbf{information gain} to form the free energy for this hypothesised state
\end{itemize}

\subsection{Information gain}
\begin{figure}
\centering
\includegraphics[width=0.7\linewidth]{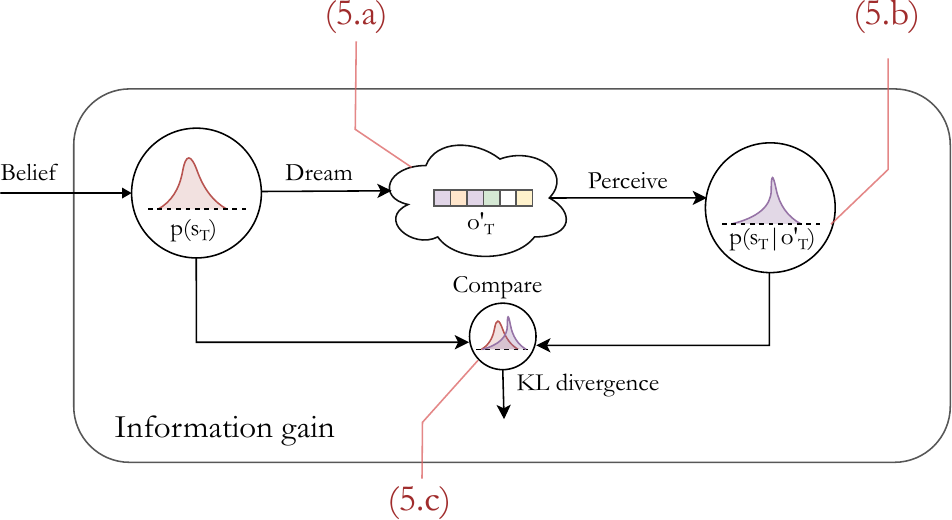}
\caption{The information gain for a future hypothesised state is computed by synthesising an observation that might be observed, performing a Bayesian update (exactly as the agent would if it really encountered this observation) and then computing how much the distribution over states has changed.
}\label{fig:ainf_inf_gain}
\Description{A diagram of the Information gain computation process}
\end{figure}

Information gain is a measure of how much \textit{could} be learned from observations in future hypothesised states. This obviously depends on a model that can synthesise plausible observations from state distributions. This generative approach is a distinctive element of Active Inference. The information gain can be computed by synthesising an observation (Fig. \ref{fig:ainf_inf_gain} (5.a)), performing a Bayesian update (Fig. \ref{fig:ainf_inf_gain} (5.b)) and then comparing the observation-updated state distribution to the original state distribution (Fig. \ref{fig:ainf_inf_gain} (5.c)). This process can then be iterated to reliably estimate the expectation of the information gain.\footnote{In some cases the (expected information gain of the) observation distribution can be computed analytically without sampling.}

\section{Mathematical definition}\label{sec:maths}
Formally, we assume an Active Inference agent interacts with an environment in discrete timesteps that we denote by index $i$. In each timestep, the environment is in a particular unobservable state $s_i \in \mathcal{S}$ and the agent interacts with its environment by choosing an action $a_i \in \mathcal{A}$. The effect of action $a_i$ on the environment is modeled by a probability distribution over successor states $P(s_{i+1} \,|\, s_i, a_i)$ commonly referred to as the \emph{transition model}. The new environment state can only be observed by the agent via a sensor state $o_i \in {O}$. The relationship between unobservable states $s_i$ and observations $o_i$ is modeled by a conditional probability distribution $P(o_i\,|\,s_i)$, commonly referred to as the \emph{observation model} or \emph{emission distribution}. The Active Inference agent holds a belief $Q_i(s)$, i.e. a probability distribution over plausible states of the environment, which is initialised from a prior belief $Q_0$ and updated using Bayesian filtering. Upon selecting an action $a_i$, the agent predicts the next environment state by updating its belief using its transition model as in Equation \eqref{eq:belief-through-time}. Upon making a new observation $o_i$, the agent corrects its prediction using its observation model as in Equations \eqref{eq:belief-after-observing} and \eqref{eq:correct-normalisation}.

\begin{align}
Q(s_{i}) &= \int{P(s_{i} \, | \, s_{i-1}, a_{i-1}) \, Q(s_{i-1}|o_{i-1})} \, \text{d}s_{i-1} &&\text{predict}  \label{eq:belief-through-time}\\
Q(s_{i}|o_{i}) &= \frac{P(o_{i} \, | \, s_{i}) \, Q(s_{i})}{P(o_{i})} &&\text{correct}\label{eq:belief-after-observing}\\
P(o_{i}) &= \int{P(o_{i} \, | \, s_{i}) \, Q(s_{i}) \, \text{d}s_{i}}&&\text{normalising constant} \label{eq:correct-normalisation}
\end{align}

The Active Inference agent's behavior is informed by a potentially time-varying prior preference distribution $P^c_i(o_i)$ that characterises the distribution over sensor states the agent would prefer to observe. The agent selects actions by approximating the expected free energy $G(\pi)$ over a range of alternative sequences of actions $\pi:(a_{i+1}, \dots a_{i+T})$ with some fixed time horizon $T$, referred to as policies, as in Equation \eqref{eq:expected_free_energy}. Note that $Q(s_k)$ and $Q(s_k|o_k)$ are hypothetical beliefs the agent anticipates to hold in the future, imagining that it followed the sequence of actions in $\pi$, that the environment follows the agent's internal transition model $P(s_{i+1} \,|\, s_i, a_i)$, and that the environment will be observed according to the agent's internal observation model $P(o_i\,|\,s_i)$. Because these beliefs are inferred from imagined futures, the correction step \eqref{eq:belief-after-observing} is skipped when constructing the sequence of hypothetical beliefs $Q(s_k)$. This step is instead only used to estimate the expected information gain. The agent defines a probability distribution $P(\pi)$ over policies using the softmax over negative expected free energy \eqref{eq:policy-softmax}, samples a policy from this distribution, and executes the first action of the sampled policy. It is sometimes convenient to define preferences $P^c(s_k)$ over latent states of the environment instead of sensor states, particularly when some aspects of the environment's state space are known and represented in the transition model. In this case, the expected free energy can be approximated using Eq.~\eqref{eq:efe_state_preference}.

\begin{align} 
G(\pi) &\approx \frac{1}{T}\sum_{k=i+1}^{i+T} -\underbrace{\E_{P(o_k\,|\,s_k)Q(s_k)}\left[ \DKL \left( Q \left( s_k | o_k \right) \middle\| Q\left(s_k\right) \right) \right]}_{\text{Information gain}} - \underbrace{\E_{P(o_k\,|\,s_k)Q(s_k)}\left[ \ln P^c_k(o_k) \right]}_{\text{Pragmatic value}} && \text{expected free energy} \label{eq:expected_free_energy}\\
&\approx \frac{1}{T}\sum_{k=i+1}^{i+T} \underbrace{\E_{Q\left(s_k\right)}\left[ \DKL\left(Q\left(s_k\right) \middle\| P^c_k\left(s_k\right)\right) \right]}_{\text{Risk}}\underbrace{-\E_{Q(s_k)}\left[ \ln P\left(o_k\,|\,s_k\right) \right]}_{\text{Ambiguity}}\label{eq:efe_state_preference}\\
P(\pi) &= \frac{\exp{\left(-G(\pi)\right)})}{\Sigma_{\pi'}{\exp{\left(-G(\pi')\right)}}} &&\text{policy distribution} \label{eq:policy-softmax}
\end{align}

The agent interacts with the environment by executing the steps above in a closed feedback loop. First, it initialises a prior belief over environment states. Second, it uses its observation from the environment to correct its belief using \eqref{eq:belief-after-observing}. Then, it selects an action using Equations \eqref{eq:expected_free_energy} and \eqref{eq:policy-softmax}. Finally, it predicts the next environment state \eqref{eq:belief-through-time}. This prediction is used as the prior belief for the next state and the cycle repeats.

\section{Vignettes}\label{sec:vignettes}
This appendix presents three short vignettes illustrating how Active Inference could be used in practical human--computer interaction problems. They are intended to make concrete some of the abstract concepts involved in Active Inference, and to highlight the challenges involved in thinking about interaction design from this perspective. 

The vignettes represent different points on the spectrum of use-cases. Semi-autonomous driving is a safety-critical application in a variable, challenging environment, with a clear task structure, emphasizing shared autonomy issues bringing sensing and skill of the driver together with the sensing and processing of the car. The soft companion robot highlights hedonic and well-being focussed interactions with little formal task structure, the importance of uncertain models of emotion, and the importance of engaging and disengaging appropriately. The intelligent music loudspeaker highlights the challenge of low-dimensional interactions with enormous amounts of data, and the use of artificial intelligence to predict human subjective responses.

We anticipate that readers may disagree with our suggestions, and in such brief descriptions of subtle tasks a brief analysis will inevitably appear superficial. However, we believe that reasoning about how an AIF agent should be designed for tasks like these can inspire new ways of thinking about the problem.
The answer to the question `What should the preferences of an AIF-based interactive agent be?' is often far from trivial, and can force us to think deeply about the design task, and potentially enable us to find new perspectives for design.


\subsection{Vignette 1: Semi-autonomous driving \texttt{U (S (U’))}}
\label{sec:driving}
\textbf{Scenario}
We imagine a vehicle with a driver support system, short of autonomous driving but with a repertoire of autonomous capabilities like cruise control, lane following, automated merging and route guidance. In this scenario, we assume that the system acts under Active Inference and has an AIF-based model of the user.

\textbf{Markov blanket}
The system can set the steering, acceleration and braking input to the vehicle. It can sense the road environment via LIDAR and the car's position and motion (via GNSS, IMUs and odometry). It can also observe the driver through the inputs from the steering wheel, brake and accelerator pedals. The system's model assumes the driver senses the environment around them (other vehicles, road markings) visually, as well as sensing acceleration forces. The system's model driver can turn the steering wheel and press the accelerate or brake pedals. We assume that other road users are merely part of the environment.\footnote{An obvious extension (section~\ref{sec:multiagentenv}) would be to model them as AIF agents \cite{engstrom2024}.}

\textbf{Forward model} 
The car needs to make several predictions: what the vehicle itself will do (e.g. turn left 5\textdegree); what the mobile elements of the environment will do (e.g. another driver approaches); and what the user will do (e.g. turn steering wheel to neutral). It further must predict how these will be sensed. If the user is concerned by a car approaching on the right, how will the control inputs change? What will its LIDAR show if the car closes distance? If the accelerator is engaged, how will the IMUs respond? As a safety-critical system a real car will require hardcoded fallback behaviours (wholly outside of the inference process) that protect the vehicle from danger. But the vehicle's predictive model should \textit{include} knowledge of those fallback behaviours so that it can model when its own control will be suppressed, just as a biological organism has awareness of its own reflexes.

\textbf{Preference prior}
The system assumes the \textit{user model} has a preference prior that (a) prefers safe and comfortable vehicle states (even, legal speed; appropriate distance to other drivers; steady turns) and (b) prefers to go to a certain destination. Without a preference to go to a destination, the car might minimise surprise entirely by refusing to leave its garage. The car's own preference prior is to align with the user's preferences, though it may have additional lower-level preferences such as to minimise mechanical wear. 

\textbf{How is surprise minimised?}
The car's autonomous function works to reduce surprise at the most basic level; e.g. its preference over a stable road speed should imply consistent measurements from odometry. Deviations (e.g. caused by changing road grade) should be planned for and acted upon to preserve the homeostasis of cruise control. It also needs to minimise surprising inputs from the driver. The driver thrashing the wheel about unexpectedly is one form of surprise and this behooves the car to carry out the implied maneuvers of the driver. For example, changing lane smoothly when the driver steers left by anticipating the control. 
The H-metaphor \cite{flemisch2003, dambock2011} proposed a model for vehicular control inspired by a horse's response to its rider that addresses how \textit{shared autonomy} could be developed. When control is definitive and meaningful, the horse does as it is bidden. When there is no meaningful control, the horse maintains comfortable homeostasis, keeping to sensible routes and returning to known locations. In much the same way, an AIF-based vehicle will execute commands compatible with its model but will autonomously act once its predictions indicate a lack of agency on the part of the driver; conversely, an increase in agency expressed through definitive but unpredicted control inputs will lead to a natural reduction of driver support.

The pragmatic values (the value derived from the preference prior) cause the car to be attracted to routes that will take it to what it believes the driver's destination to be. Driver inputs to make navigational changes counter to the car's route planning should be minimised, so the car should not only take a route that leads them to their destination, but one that the car believes the driver will expect to be taken -- and not a clever detour through an industrial area.

\textbf{Why is Active Inference appropriate?}
Driving is clearly a task where predictive models are vital. Driving involves predictions on the scale of milliseconds to predict vehicle dynamics, on the scale of seconds for how other road users will behave, and on the scale of minutes for how route decisions will impact navigation. Likewise, uncertainty is clearly relevant. Vehicle dynamics are relatively predictable but other drivers are not. Minimising surprise at the vehicle level has obvious correlates with safe driving behaviour, but the reflective component of surprise minimisation is particularly interesting here. Autonomous driving supplants control that the driver used to have to provide and taking on this workload without constraining what the driver can do is a challenge in shared autonomy. Reflective surprise minimisation supports this handover.
The vehicle plans to minimise the surprise from driver controls, which it predicts based on a user model that wants to minimise surprise about what they in turn observe, given their intentions. This is intrinsic motivation for the car to establish what the driver's intentions are and to follow them correctly \textit{and visibly}. When the driver's inputs \textit{are} poorly predicted, one way for the car to minimise future surprise is to temporarily propagate inputs more directly; ``tightening the reins'' in the H-metaphor sense, and making the driver more responsible for direct control.

\subsection{Vignette 2: A soft companion robot \texttt{U (S)}}
\textbf{Scenario}
We imagine an artificial companion fulfilling roughly the role of a domestic cat, for example supporting older populations who cannot care for a live animal \cite{WadaShibata2007}. 

\textbf{Markov blanket} Physically, the imaginary device is a robot with a pliable exoskeleton, limited but high-degree-of-freedom actuation that can flex its soft exterior, and a range of surface-based touch sensors distributed over its body. 

\textbf{Preference prior} What does it prefer? It prefers its companion human to be happy and relaxed.  It might have a dimensional model of its companion's emotional state (e.g. an arousal-valence model \cite{russell1980}).\footnote {While modelling emotion might appear challenging, AIF approaches have made significant strides in computational modelling emotional responses. \cite{pattisapu2024free,barrett2017k}} Its preference prior would then be close to zero for emotional coordinates representing sadness, anger, agitation, etc. 

\textbf{Forward model} The robot makes predictions over the mental state of the human it supports, driven by a forward model that relates sensing and actuation to emotional responses. This in turn depends on a predictive model of how its movements affect its own sensors; for example, a physical model of contact that can predict that flexing against an opposing surface will increase activity at those touch sensors pressed against the surface. 

\textbf{How is surprise minimised?} The robot minimises surprise, where it is surprised by having an unhappy human. Its actions will be motivated by a combination of keeping the user in preferred mental states and monitoring what those are. As a top-level behavior, it may act to stabilise what it believes to be a comforting state ("keep moving gently"); it may act to stimulate a bored or disengaged user ("move in an amusing way") or back off from a tired or irritated user ("go to sleep"). But it may also act to elicit how the companion feels, or even if their companion is present -- ("squirm and see if something happens").

At a lower level, it needs to act to make the most of its own body. It might contort itself such that its sensory apparatus -- its active skin -- is exposed to touch to engage with the user (minimising surprise through inquisitive kinematics). It may also act to establish to what degree it can move ("am I being hugged?") by actuating its flexors and seeking change at its sensors. It must predict both how these self-calibrating actions will inform the agent of its immediate environment \textit{and} how those actions will be perceived by the user. Extreme flexes are likely to distress the user, even if they are highly informative for the robot.

\textbf{Why is Active Inference appropriate?} Surprise minimisation naturally drives a trade-off between soothing the companion's emotional states, establishing what they are, and optimising how they can be sensed and acted on. The automatic balance between inquisitiveness and utility maximisation at multiple levels is a notable virtue of Active Inference that would be hard to replicate otherwise. Any model of a human's emotional response will be incomplete. Humans are complex, and the robot's sensorium has limited engagement with its companion's internal states. 
This makes it vital to adequately represent uncertainty. It also motivates predictive models, since the lag between movement and emotional response is substantial. A purely reactive system would be unstable and difficult to engineer. 

\subsection{Vignette 3: An intelligent music loudspeaker \texttt{U (S (U'))}}

\textbf{Scenario}
This vignette presents a hypothetical AIF-powered consumer device, a smart music speaker.  Its actions are to change the music played and a small two-dimensional display of a `music map' which presents an abstract display of the space of all possible musical tracks. Its sensorium (the totality of its sensing) comprises a room-level motion sensor, an internal clock and a radar proximity sensor of the style described in Figure~\ref{fig:hand_radar}, which can detect the distance and velocity of the user's hand in the 20cm directly above the device. It can also detect whether the user's hand has spread fingers or not.

\textbf{User satisfaction represented via preferences/goals: }
The user can be assumed to have long-term musical preferences (we will assume the system has an uncertain and incomplete knowledge of these), but might be affected by their immediate context, e.g. they may also play music for social reasons,\footnote{\textbf{Environment: }If a user were listening to music on their own, with headphones on, the environment would be limited in its impact on the system. However, if the user were controlling the music to change the mood of a party of other people in their house, the nature of the task is different, leading to different effective preferences from the single use case, and the user's attention is divided between the interface and the broader audience.} and might have other short-term goals such as learning more about a particular genre, and hence their preferences can adapt over time. Preferences might also include having no music at all (zero volume) in certain contexts (e.g. tired, needing to work, social context).

\textbf{How is surprise minimised?}
The speaker's apparent purpose is to play music, but a more fundamental purpose is to keep its user satisfied. As an AIF agent, the {\it speaker} achieves its purpose by adjusting the state in music space, and hence changing the intermediate visualisation and the music played. By affecting the user in an appropriate way, this makes the system's {\it future} sensory predictions of its motion and proximity sensors as accurate as possible.
This is the principle of surprise minimisation. It cannot perceive satisfaction directly, but only via the effect of its actions as reflected through the lens of the user's sensed actions.  If we assume that the {\it user} behaves like an AIF system their actions will be to behave in such a way that the system responds predictably to their actions and in a manner aligned with their preferences.

So let us examine how this mutual prediction could be realised in practice, based on user actions and feedback: \\
\noindent \textbf{User Actions: }The music system needs to be able to take low-dimensional inputs from the user and use these to navigate a high-dimensional space of music tracks, akin to the semi-autonomous driving example from the previous section. However, in this case there are not enough degrees of freedom or accuracy in the input to select arbitrary tracks, at the resolution needed, from sensed absolute position or hand spread alone, so the extra information must come from the interaction between the human and the music system.\footnote{There are many possible ways to envisage the implementation such a system (e.g. learning low-dimensional projections of the high-dimensional space, and combining these with local searches, based on sojourn time in an area), but for any given candidate system, we should be able to act in such a way that we minimise the EFE, where actions can include learning parameterisations of embeddings or local dynamics. }

\subsubsection{Forward models}
To enable interaction requires multiple levels of feedback loops, and each of these loops will have associated predictive forward models.
Prediction: We need to learn four groups of forward probabilistic models. We need to (1) be able to predict mappings from human action to the speaker's sensor, and hence to the state of the input device. We then need (2) to map from that state to the change in music space state, (3) from that state to the human subjective music experience and (4) from the subjective experience of the user to a user action.

\noindent \textbf{1. System Forward Model -- sensing:}  We need forward models of the sensor system, in order to infer the user's actual physical movement, as described in \ref{sec:forward}, so that the basic low-dimensional input can work as expected by the user. Hand proximity 
is used to control the state of the input device, and the finger spread can  indicate uncertainty or vagueness.

\noindent \textbf{2. User Forward Model -- intermediate display:}
The state of the low-dimensional input device will be fed into the music state space transition dynamics, and will change the state in music space in a deterministic manner. The user model will have a probabilistic forward model which attempts to predict the change in this display given the state of the input device (which is driven by their low-dimensional actions). 

\noindent \textbf{3. User Forward Model -- human subjective music experience:} The features of the music space correspond to an implicit model of how the user will subjectively experience those tracks. Particular intermediate display settings should lead to a predictable musical experience for the user. \footnote{This assumes that the intermediate display setting corresponds to where the music is actually selected. In low-density areas, however, there may be no tracks nearby, so we may end up playing inaccurate content, and surprising the user, leading to a poor free energy measure. A further subtlety is that the user's ability to differentiate different classes of content may vary, in which case deviations between the intermediate display setting and played content might not be noticed by one user, but would be by others. This can be represented by uncertainty in the AIF model of the user.}

\noindent \textbf{4. User Forward Model -- human control of music space:} Once the human has perceived the nature of the current track playing, they may be satisfied, and not make further action, or might want to change, and must decide where to move to in music space. This will require the user to be able to predict where in the music space they would be happier, and use User Forward Model 2 to predict what actions would take them there. The user action model will also have the user's own approximate model of System Forward Model 1 which describes the reliability of action sensing.

\subsubsection{Designing a system}

\textbf{Incorporating Machine learning methods:}
We now assume that machine learning can be used to learn offline from training sets of human-labelled data, and create an ordinal regression model which can analyse the audio tracks and probabilistically predict human subjective responses for multiple features such as mood, genre, tempo etc. This can be used to pre-index hundreds of thousands of tracks with these subjective responses in advance. This enables the creation of a 2D embedding of the multidimensional music space to visualise the location of any of the tracks. This simple visual display shows a summary of responses to human sensed motion, and the projection of the current location in `music space' which is intended to help provide the user with a bridge to understand how the music is going to be generated. This could be a coloured two-dimensional `music map', as in \cite{VadBolWil15AI}. From an individual point in the music space, the system can sample the nearest $N_t$ points from this location, and generate playlists. 
It has a sense of time, and can sense when the user is in the room, so can initiate music playing, when appropriate.


\subsubsection{Preference alignment} 
The previous section described how the free energy approach would support making an interface predictable, but does not yet take user preferences into account. If the system side had an internal AIF model of the user, including learned preference priors for musical taste, then the system could combine the density of content with the preference weighting to move to regions of musical content the user most likes to listen to. This might be experienced by the user as them controlling one aspect of the input, and the music system responding proactively with a sequence of `offers' of movement, which it could accept or reject via subtle movements of the input. We give illustrations of interaction in section~\ref{sec:interacthist}.

\subsubsection{Illustrative interaction histories}
\label{sec:interacthist}
In Section~\ref{sec:principles} we observed that Active Inference {\it `unifies perception and action by considering agents as if they stood at the end of predicted paths into all possible futures and looked back to infer which actions would be most likely to have taken along their preferred paths'}, so let us imagine several cases.


\subsubsection*{Paths to preferred and non-preferred tracks}
If we assume that the smart speaker's predictive model includes an approximate model of the user's music preferences, how would the interaction history evolve differently for the two cases of a user with a style of music in mind wishing to select content which has 1. a high preference versus 2. a low preference?

In this case, we assume an experienced user who understands the genre/mood mappings implicit in the 2D map. In case 1., if they are aiming for a region with their preferred content then a rapid movement to the approximate area, although being less informative because of its higher error rate, may still be sufficient to rapidly select preferred content, as even if they do not get to the ideal region after the first action, the interface is more likely to offer preferred content nearby, so with a small number of interactions they will enter an area with their desired content and stop interacting to listen. The overall effect is similar to that observed in Fitts' Law tasks, where users move more rapidly to larger targets because of the greater leeway for error.

However, if, as in case 2, the target track is unusual for them, the user will predict that they will get little useful support from the system to find it, as it will initially assume more preferred content, and they will need to generate a flow of precise (and hence informative) movements. As with smaller spatial targets in a traditional interface, they will make more careful, precise movements, with more time to observe where they are in the music space, or listen to the content for feedback about how to change their position with their next action. The system can observe this unusual behaviour, and infer that it is compatible with a user acting carefully because they are aiming for a less likely target, and it can increase the uncertainty on the preference structure, making less preferred tracks more likely to be presented. 

\subsubsection{Why less is often more, with Active Inference}
When comparing possible paths to the goal, the system will tend to see as less likely paths which generate more uncertainty than required. For example:
If the system knows that the user's perceptual model is less able to distinguish certain genres, it will be less likely to offer contrasts between those genres, as any responses will be less informative than between genres/moods that the user can more reliably separate. Similarly the user will be less likely to choose to navigate through such regions where they have poor acuity, if routes through other areas are more predictable for them. 

The timing and execution of user interactions in general is variable, meaning that the sensations at its sensor are inherently unpredictable; and they represent a breach of the agent's comfortable homeostasis. So if a user does interact with the speaker the system will be motivated (by its intrinsic goal to minimise surprise) to make the interactions as short and simple as possible, and entrain the user to move predictably, to reduce its sensory surprise. It also becomes more likely to act pre-emptively when it predicts a user intervention.

Although its only \textit{external action} is to adjust the music played, and its display, internally it can build models of what percept influences changes in behaviour and update its model to improve its predictions. If it detects the user moving out of range, it may learn to turn down the sound to avoid an unpredicted touch. It may learn activity patterns, so that certain movements -- say, typing or working out -- can be recognised and responded to. It may learn a temporal model, adjusting the volume and type of music as evening progresses, if it has had to endure frequent manual adjustments in the past. If its touch sensor is poorly calibrated to the user's expectations, they may spend time fiddling with it. And so it has an implicit objective to calibrate its touch to suit the user, to minimise these interactions. If its predictions become weak it may actively elicit a response to reinforce or contradict them. For example, if the motion sensor reads nothing, it may be ambiguous as to whether the user is still or has departed; varying the audio slightly will provoke a response from the user that will quickly disambiguate the uncertainty.




\subsubsection{Why is Active Inference appropriate?}
This approach to design is a novel one. The speaker is designed to minimise surprise. From this simple principle we can envision intelligent, adaptive behaviour where many of the ways it ``should work'' are not manually designed but arise naturally from the free energy principle. This responsive behaviour adapts in the moment by continuously planning over anticipated sensation; it is not a collection of pre-baked responses. Proactive, dynamic interaction can bridge the gap between the low-dimensional input and the high dimensional content space.

Active Inference can also provide a clear structure for using machine learned models of human subjective experience of content, and associated preferences. This means that models of user preference, of user search and of foraging behaviour can be cleanly included to inform and improve the interaction.

\end{document}